\documentclass[12pt]{article}
\usepackage{graphicx,latexsym}

 \def\dag{\dagger} \def\pd{\partial} \def\pp{\prime} \def\a{\alpha} \def\b{\beta} \def\dl{\delta} \def\s{\sigma}   \def\eps{\epsilon} 
\def\veps{\varepsilon} \def\lam{\lambda} \def\Lam{\Lambda} \def\gm{\gamma} \def\Gm{\Gamma}
\def\om{\omega} \def\Om{\Omega} \def\nb{\nabla} \def\sq{\sqrt} \def\e{\hbox{\large \it e}}
\def\half{\frac{1}{2}} \def\fr{\frac} \def\arr{\rightarrow} \def\prm{m^\prime}

 \def\QG{{\rm QG}} \def\pl{{\rm pl}} \def\M{{\rm M}} \def\T{{\rm T}} \def\S{{\rm S}} \def\R{{\rm R}}     
 \def\L{{\rm L}}  \def\V3{{\rm V}_3}

 \def\hg{{\hat g}} \def\hDelta{{\hat \Delta}} \def\hnb{{\hat \nabla}} \def\hgm{{\hat \gamma}} \def\hR{{\hat R}} \def\hE{{\hat E}}

     \def\C{{\bf C}}
\def\D{{\bf D}} \def\E{{\bf E}}  \def\H{{\bf H}}

\def\lap3{~| \!\!\! \partial^2} \def\dlap3{~| \!\!\! \partial^4}

\def\lang{\langle} \def\rang{\rangle}

\def\l:{: \!}\def\r:{\! :}

\def\h{{\sf h}}     
\def\rb{{\rm b}} \def\rc{{\rm c}} \def\gh{{\rm gh}}  \def\BRST{{\rm BRST}}

 \def\rvac{|0 \rangle}

\begin{document}

\begin{titlepage}

\begin{flushright}
July 2017
\end{flushright}

\vspace{5mm}

\begin{center}
{\Large {\bf BRST Conformal Symmetry as A Background-Free Nature of Quantum Gravity}}
\end{center}

\vspace{5mm}

\begin{center}
{\sc Ken-ji Hamada}
\end{center}

\begin{center}
{\it Institute of Particle and Nuclear Studies, KEK, Tsukuba 305-0801, Japan} \\ and \\
{\it Department of Particle and Nuclear Physics, SOKENDAI (The Graduate University for Advanced Studies), Tsukuba 305-0801, Japan}
\end{center}

\begin{abstract}
Quantum gravity that describes the world beyond the Planck scale should be formulated in a background-metric independent manner. Such a background-free nature can be represented as a gauge equivalency under conformal transformations, called the BRST conformal symmetry. In this review, we present quantum field theories of gravity with such symmetry. Since we can choose any background owing to this symmetry as far as it is conformally flat, we here employ the cylindrical background. First, we briefly review the famous BRST Liouville-Virasoro algebra in 2D quantum gravity on $R \times S^1$. We then present recent developments of the BRST conformal algebra and physical states of 4D quantum gravity on $R \times S^3$ whose conformal-factor dynamics is ruled by the Riegert's Wess-Zumino action, which arises in the UV limit of the renormalizable quantum conformal gravity with the ``asymptotic background freedom". We find that the BRST conformal invariance makes all physical states real and scalar as well as all negative-metric modes unphysical.

We also briefly discuss the dynamics of how the conformal invariance breaks down and how our classical spacetime emerges at low energies, in which a novel interpretation of the ``minimal length" we can measure comes out without discretizing spacetime. 
\end{abstract}

\vspace{5mm}

\end{titlepage}

\tableofcontents

\section{Introduction}
\setcounter{equation}{0}

It is well-known that the Einstein gravity is perturbatively unitary, but non-perturbatively it is not well-defined because the Einstein-Hilbert action is not bounded from below and there are spacetime singularities such as the Schwarzschild black hole. Since the action becomes finite even for such singular solutions, singularities cannot be removed within the Einstein gravity. As long as this singularity problem is not overcome, we cannot look into the world beyond the Planck scale.

Beyond the Planck scale, in the first place, the conventional picture that a point-like elementary particle is propagating in the flat background breaks down within the framework of the Einstein gravity. It is because if the particle mass $m$ is greater than the Planck mass $m_\pl$, its Compton wave length becomes less than its horizon size and thus information of such a particle will be hidden inside the horizon and lost.

\begin{figure}[h]
\begin{center}
\includegraphics[scale=0.7]{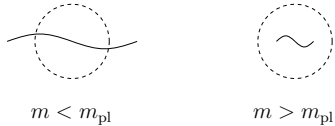}
\end{center}
\vspace{-8mm}
\caption{\label{particle picture violation}{\footnotesize The typical size of a particle with the mass $m$ is given by the Compton wave length $\lam \sim 1/m$. On the other hand, its horizon size (dotted line) is given by $r_g \sim m/m_\pl^2$. For $m > m_\pl$, $\lam < r_g$, and thus information of such an elementary excitation will be lost.}}
\end{figure}

In order to resolve the singularity problem as well as renormalizability, quantum gravity theories with higher-derivative actions have been studied in the 1970s \cite{stelle, tomboulis, ft, ab}. The positive-definite action including the square of the Riemann curvature tensor becomes well-defined non-perturbatively, because for a singular spacetime solution this action diverges and thus such a spacetime is forbidden quantum mechanically.

In higher-derivative theories, however, the unitarity problem arises, which is that the graviton mode with negative metric emerges as a gauge-invariant state. This problem becomes manifest in the UV limit of the conventional perturbation theory based on the asymptotically free behavior.\footnote{ 
It is known that there is an idea by Lee and Wick to avoid the ghost pole by including the running coupling effect \cite{lw, nakanishi, bg}, but such an effect disappears in the UV limit.} It also suggests that the particle picture propagating on the flat background is not feasible beyond the Planck scale.

In order to resolve the unitarity problem, it seems necessary to incorporate a non-perturbative approach. For several years ago, we proposed the renormalizable quantum gravity theory with the background-free property, that is represented by the BRST conformal symmetry, in the UV limit  \cite{hs, hamada02, hamada14b, hm16, hm17}. This symmetry is realized by treating the conformal factor of the metric field in a non-perturbative manner \cite{riegert, am, amm92, amm97a, amm97b, hh, hamada05, hamada09, hamada12a, hamada12b}. It is a gauge symmetry that arises as a part of diffeomorphism invariance, and thus it represents that all theories connected one another by conformal transformations become gauge-equivalent, as a realization of the background-metric independence. It resolves the unitarity problem by making negative-metric modes unphysical. This behavior is called the ``asymptotic background freedom". The particle picture is then abandoned.

In general, the BRST conformal symmetry can be defined in even dimensions. In order to define quantum gravity with this symmetry, we decompose the metric field as follows:
\begin{eqnarray}
     g_{\mu\nu}   =  e^{2\phi} \left( \hg e^{th} \right)_{\mu\nu} 
         = e^{2\phi} \hg_{\mu\lam} \left( \dl^\lam_{\nu}  +  t h^\lam_{\nu}  +  \cdots \right),
       \label{metric decomposition}
\end{eqnarray}
where $\hg_{\mu\nu}$ is the background metric and the traceless tensor field satisfies $h_{\mu\nu}=\hg_{\mu\lam} h^\lam_{~\nu}$ and $h^\mu_{~\mu}=0$. The conformal-factor field $\phi$ is quantized non-perturbatively without introducing its own coupling constant, while the traceless tensor field $h_{\mu\nu}$ is handled by the perturbation theory in $t$. In the following, the quantity with the hat denotes that it is defined on the background metric.

We here consider quantum gravity that has a conformally invariant action $I$ in the UV limit where the terms with mass parameters can be disregarded. In 2 dimensions, $h_{\mu\nu}$ has no dynamical degrees of freedom and thus, apart from the topological term, $I$ is a matter-field action only, which is described by a conformal field theory (CFT). On the other hand, in 4 dimensions, $h_{\mu\nu}$ is a dynamical field, which is managed by the Weyl action divided by $t^2$. We here consider the $t \to 0$ limit so that only the kinetic term of the traceless tensor field remains. This limit is given by the UV limit above the Planck mass scale and there the background-free dynamics is realized. Thus, the coupling $t$ indicates the asymptotically background-free behavior. It denotes a deviation from the background-free system and so its running coupling will connect between background-free world and our present background-dependent world.

Quantum gravity is defined by the path integral over the metric field weighted by the action $I$. Techniques to treat diffeomorphism invariance are first developed in 2 dimensions \cite{polyakov, kpz,dk, seiberg, teschner, gl, bmp}, and extended later in 4 dimensions \cite{riegert, am, amm92, amm97a, amm97b, hs, hh, hamada05, hamada09, hamada12a, hamada12b}. It is that we change the path-integral measure from the diffeomorphism invariant measure to the practical measure defined on the background as follows:
\begin{eqnarray}
   Z =\int [dg df]_g e^{iI(f,g)} = \int [d\phi d h df]_\hg  e^{iS(\phi,\hg)+ iI(f,g)} ,
                  \label{path integral expression}
\end{eqnarray}
where $f$ is a matter field. In the second equality, the Wess-Zumino action $S$ \cite{wz} associated with conformal anomalies \cite{cd, ddi, duff, hathrellS, hathrellQED, freeman, ds, hamada01, hamada14a} is necessary as Jacobian to recover the diffeomorphism invariance. It satisfies the consistency relation  
\begin{eqnarray} 
     S(\phi-\om,e^{2\om}\hg)+S(\om,\hg) = S(\phi,\hg) .
        \label{Wess-Zumino relation for action S}
\end{eqnarray}
In 2 dimensions, it is the famous Liouville action \cite{polyakov}, and the 4-dimensional one is given by the Riegert action \cite{riegert}. As defined later, they are obtained by integrating the respective conformal anomalies by the conformal-factor field.

Here, note that conformal anomalies violate classical conformal invariance in quantum field theories on curved spacetime. When quantizing gravity, however, conformal anomalies play a significant role to recover exact conformal invariance, namely background-metric independence, because conformal anomalies are physically not anomalous quantities, which in general arise to ensure diffeomorphism invariance, as mentioned below.

Using the Wess-Zumino consistency relation (\ref{Wess-Zumino relation for action S}), we can show the background-free nature of the theory as follows:\footnote{ 
The overall coefficient of the Wess-Zumino action $S(\phi,\hg)$ is given by the coefficient of the conformal anomaly, which is given by (\ref{coefficient of Liouville action}) and (\ref{coefficient of Riegert action}) in each dimension. It includes a contribution from the Wess-Zumino action itself, and thus when rewriting the measure $[dgdf]_g$ with $[d\phi dhdf]_\hg e^{S(\phi,\hg)}$ in (\ref{path integral expression}), it has exactly to be determined in a self-consistent manner such that the theory becomes background free. It is that since the Jacobian factor $S(\om,\hg)$ can be evaluated using the action $S(\phi,\hg)$ itself, independently of its coefficient, we can set the coefficient of $S(\phi,\hg)$ to that of the calculated $S(\om,\hg)$ so as to be background free as in (\ref{background free property}).}
\begin{eqnarray}
   Z(e^{2\om}\hg) &=&  \int  [d\phi dh df]_{e^{2\om}\hg} 
     e^{ iS(\phi,e^{2\om}\hg)+iI(f,e^{2\om}g) }
              \nonumber \\
    &=& \int [d\phi dh df]_\hg e^{iS(\om,\hg)} e^{iS(\phi,e^{2\om}\hg)+iI(f,g) }
              \nonumber \\
    &=& \int [d\phi dh df]_{\hg} \, e^{iS(\om,\hg)+ iS(\phi-\om,e^{2\om}\hg)+iI(f,g) } = Z(\hg) .
       \label{background free property}
\end{eqnarray}
Here, in the second equality, we take into account the Jacobian factor $\exp\{iS(\om,\hg)\}$ arising when we rewrite the path integral measure. The third equality is derived by changing the variable of integration as $\phi \to \phi-\om$, considering that the measure $[d\phi]_\hg$ is invariant under such a local shift. In the last equality, we use the consistency relation (\ref{Wess-Zumino relation for action S}).

The BRST conformal invariance we discussed here is just an algebraically rigorous representation of this background-metric independence. It arises as a part of diffeomorphism invariance as follows. The diffeomorphism invariance is defined by the transformation $\dl_\xi g_{\mu\nu}=g_{\mu\lam}\nb_\nu \xi^\lam + g_{\nu\lam}\nb_\mu \xi^\lam$, which is now expanded as
\begin{eqnarray*}
   \dl_\xi \phi  &=& \xi^\lam \pd_\lam \phi + \fr{1}{D} \hnb_\lam \xi^\lam ,     
               \nonumber \\
   \dl_\xi h_{\mu\nu}  &=&  \fr{1}{t} \left( \hnb_\mu \xi_\nu  +  \hnb_\nu \xi_\mu 
           -  \fr{2}{D} \hg_{\mu\nu} \hnb_\lam \xi^\lam \right)
           +   \xi^\lam \hnb_\lam h_{\mu\nu}   
           +  \fr{1}{2} h_{\mu\lam}  \left(   \hnb_\nu \xi^\lam   -  \hnb^\lam \xi_\nu   \right) 
                 \nonumber \\
      &&   +  \fr{1}{2} h_{\nu\lam}  \left(   \hnb_\mu \xi^\lam   -  \hnb^\lam \xi_\mu   \right) 
           + o(t)  ,
\end{eqnarray*}
where $\xi_\mu = \hg_{\mu\nu} \xi^\nu$ and $D$ is the spacetime dimensions. The conformal symmetry arises when the gauge parameter $\xi^\mu$ is given by the conformal Killing vectors $\zeta^\mu$ satisfying $\hnb_\mu \zeta_\nu + \hnb_\nu \zeta_\mu - 2 \hg_{\mu\nu} \hnb_\lam \zeta^\lam/D=0$.  At $t \to 0$, we obtain the conformal transformation
\begin{eqnarray}
    \dl_\zeta \phi  &=&  \zeta^\lam \pd_\lam \phi + \fr{1}{D} \hnb_\lam \zeta^\lam ,
              \nonumber \\
    \dl_\zeta h_{\mu\nu}  &=& \zeta^\lam \hnb_\lam h_{\mu\nu} 
         + \half h_{\mu\lam} \left( \hnb_\nu \zeta^\lam - \hnb^\lam \zeta_\nu \right)
         + \half h_{\nu\lam} \left( \hnb_\mu \zeta^\lam - \hnb^\lam \zeta_\mu \right)  \quad
       \label{conformal transformations with zeta^mu}
\end{eqnarray}
as a gauge transformation.\footnote{ 
Both $\phi$ and $h_{\mu\nu}$ transform as dimensionless fields. Since they are gauge fields, they themselves do not necessary to satisfy the unitarity bound for primary fields, just like vector gauge fields are so.} 
Note that the right-hand side of this gauge transformation is field-dependent, and thus all modes in the fields are mixed by gauge symmetry even when the interaction is turned off. It means that each mode itself is not gauge invariant even in the UV limit.

The BRST conformal transformation is defined by replacing the gauge parameter $\zeta^\mu$ with the corresponding ghost field $c^\mu$. In the following, we will consider the quantum gravity systems that the traceless tensor field is gauge-fixed properly so as to reduce the residual gauge degrees of freedom to the conformal Killing vectors only.

In this paper, we review the quantum symmetry of (\ref{conformal transformations with zeta^mu}) in $2$ and $4$ dimensions. Due to the BRST conformal invariance, we can employ any background as long as it is conformally flat. Here, we choose the cylindrical background $R \times S^1$ and $R \times S^3$.\footnote{ 
The formulation employing the Minkowski background $M^4$ is given in \cite{hamada12a}.} 
Our conventions are the following: the signature of the time component of the metric is negative and the curvatures are defined by $R^\lam_{~\mu\s\nu}=\pd_\s \Gm^\lam_{\mu\nu}- \cdots$ and $R_{\mu\nu}=R^\lam_{~\mu\lam\nu}$.

\section{2D Quantum Gravity on $R \times S^1$}
\setcounter{equation}{0}

To begin with, we briefly review 2D quantum gravity as the simplest model with the BRST conformal symmetry. This type of the BRST symmetry has been first formulated in string theory \cite{kato, fms} and then developed into 2D quantum gravity \cite{lz,bmp}.

The traceless tensor fields have 2 degrees of freedom in 2 dimensions. They can be, however, completely gauge-fixed using 2 gauge degrees of freedom $\xi^\mu$. So, we here choose the conformal gauge condition defined by
\begin{eqnarray}
  h_{\mu\nu}=0 .
  \label{conformal gauge fixing}
\end{eqnarray}

In this gauge, the action of 2D quantum gravity is given by $S_{\rm 2DQG} = S_\L +I_\M + I_{\rm gh}$, where $I_\M$ is the action of the matter CFT and $I_{\rm gh}$ is the $bc$-ghost action in the conformal gauge defined later. $S_\L$ is the Liouville action \cite{polyakov} defined by\footnote{ 
Usually, the field is rescaled as $\phi \to \phi/\sq{2b_\L}$ and the action is rewritten in the form $-(1/8\pi)\{ (\pd \phi)^2 + Q \hR \phi\}$, where $Q=\sq{2b_\L}$. But here, to emphasize the resemblance to 4D quantum gravity later, we quantize it without rescaled.} 
\begin{eqnarray*}
    S_\L &=& - \fr{b_\L}{4\pi} \int d^2 x \int^\phi_0 d\phi \sq{-g} R
                  \nonumber \\
         &=& - \fr{b_\L}{4\pi} \int d^2 x \sq{-\hg} \left( \phi \hDelta_2 \phi + \hR \phi \right) ,
\end{eqnarray*}
where $\sq{-g}\Delta_2 = \sq{-g}(-\nb^2)$ is the conformally invariant differential operator for a scalar in 2 dimensions. When we perform the integration over $\phi$, the relation $\sq{-g}R=\sq{-\hg}(2\hDelta_2 \phi +\hR)$ for $g_{\mu\nu} =e^{2\phi}\hg_{\mu\nu}$ is useful. The coefficient in front of the Liouville action is given by \cite{kpz,dk}
\begin{eqnarray}
     b_\L = - \fr{c_\M - 25}{6} ,
        \label{coefficient of Liouville action}
\end{eqnarray}
where $c_\M$ is the central charge of the matter CFT and $-25$ in the numerator is the sum of $1$ and $-26$ coming form the Liouville field and $bc$-ghosts, respectively. So, for $c_\M < 25$, the Liouville action has the right sign $b_\L >0$.

\subsection{Virasoro Algebra and Physical States}

The background free property (\ref{background free property}) ensures that the theory is independent of how to choose the background metric, as far as it is conformally flat. Here, we quantize the gauge-fixed action $S_{\rm 2DQG}$ by choosing the cylindrical background $R \times S^1$ parametrized as $x^\mu=(\eta,\s)$, where $0 <\s<2\pi$.

From the equation of motion, the $\phi$ field is expanded using the left- and right-moving modes and zero-modes as
\begin{eqnarray*}
   \phi(\eta,\s) =  \fr{1}{\sq{2b_\L}} \Bigl\{ 
          \hat{q} + 2\eta \hat{p} +  \sum_{n \neq 0}\fr{i}{n} 
            \left( \a^+_n e^{-in (\eta+\s)}  + \a_n^- e^{-in(\eta-\s)} \right) \Bigr\} .
\end{eqnarray*}
Since $\phi$ is a real field, the Hermite conjugate of each mode is defined by $\a^{\pm\dag}_n =\a^\pm_{-n}$. The conjugate momentum is given by $\Pi=(b_L/2\pi) \pd_\eta \phi$ and the equal-time commutation relation is set as $[\phi(\s), \Pi(\s^\pp)]=i \dl(\s-\s^\pp)$. The commutation relation for each mode is then given by
\begin{eqnarray*}
    [\hat{q},\hat{p}]=i, \quad [\a^\pm_n, \a^\pm_m ] = n\dl_{n+m,0}, 
    \quad [\a^\pm_n, \a^\mp_m ] = 0 .
\end{eqnarray*}

In the conformal gauge (\ref{conformal gauge fixing}), there exist the residual gauge degrees of freedom $\zeta^\mu$ satisfying the conformal Killing equations 
\begin{eqnarray*}
       \pd_\mu \zeta_\nu + \pd_\nu \zeta_\mu -\eta_{\mu\nu} \pd^\lam \zeta_\lam =0 .
\end{eqnarray*}
It can be found from the fact that taking $\xi^\mu=\zeta^\mu$ we obtain $\dl_\zeta h_{\mu\nu}=0$ in the conformal gauge so that the gauge condition is preserved. Using this residual gauge degrees of freedom $\zeta^\mu$, we can construct the conformal algebra. The generator of the transformation $\dl_\zeta \phi$ in (\ref{conformal transformations with zeta^mu}) is defined by
\begin{eqnarray}
       L_{\zeta} = \int_{S^1} d\s  \zeta^\mu \l: {\hat T}_{\mu 0} \r: ,
         \label{definition of generator of Virasoro algebra}
\end{eqnarray}
where the symbol $: \,\, :$ denotes the normal ordering. The energy-momentum tensor is defined by ${\hat T}^{\mu\nu} = (2/\sq{-\hg}) \dl S_{\rm 2DQG}/\dl\hg_{\mu\nu}$.

The energy-momentum tensor of the Liouville sector is given by
\begin{eqnarray*}
   {\hat T}^\L_{\mu\nu} = \fr{b_\L}{2\pi} \left\{ 
        \pd_\mu \phi \pd_\nu \phi -\half \eta_{\mu\nu} \pd^\lam \phi \pd_\lam \phi 
        + \left( \eta_{\mu\nu} \pd^\lam \pd_\lam - \pd_\mu \pd_\nu \right) \phi 
        \right\} .
\end{eqnarray*}
Here, the last linear term is the characteristic part of the Liouville theory, which comes from the $\hR\phi$ term. The trace of the energy-momentum tensor vanishes by the equation of motion and thus the generator (\ref{definition of generator of Virasoro algebra}) is conserved.

In 2 dimensions, there are an infinite number of the conformal Killing vectors $\zeta^\mu$, which are given by $(e^{in(\eta+\s)}/2,e^{in(\eta+\s)}/2)$ and $(e^{in(\eta-\s)}/2,-e^{in(\eta-\s)}/2)$ for integer $n$. Substituting these into (\ref{definition of generator of Virasoro algebra}), we obtain the Virasoro generator
\begin{eqnarray*}
     L^{\L\pm}_n  &=& e^{in\eta}  \int_0^{2\pi}  d\s  e^{\pm in\s} 
           \half \l: \bigl( \hat{\Theta}^\L_{00} \pm \hat{\Theta}^\L_{01} \bigr) \r: + \fr{b_\L}{4} \dl_{n,0}
                 \nonumber \\
     &=&  \half \sum_{m \in {\bf Z}} \l: \a^\pm_m \a^\pm_{n-m} \r: 
           + i \sq{\fr{b_\L}{2}}n \a^\pm_n + \fr{b_\L}{4} \dl_{n,0} ,
\end{eqnarray*}
where $\a^\pm_0=\hat{p}$ and the Hermitian condition is given by $L^{\L\pm\dag}_n = L^{\L\pm}_{-n}$. We here add the last term $(b_\L/4)\dl_{n,0}$, which is the Casimir effect on $R \times S^1$. It shifts the Hamiltonian by the constant $b_\L/2$ as
\begin{eqnarray*}
    H^\L = L^{\L +}_0 + L^{\L -}_0 =\hat{p}^2+ \fr{b_\L}{2} + \sum_{n=1}^\infty \left\{
         \a^{+ \dag}_n \a^+_n +  \a^{- \dag}_n \a^-_n \right\} .
\end{eqnarray*}
This energy shift is necessary that the Virasoro algebra closes quantum mechanically.

The full Virasoro generator, including those for the matter field and the ghost field defined later, is denoted by
\begin{eqnarray*}
   L_n^\pm= L^{\L\pm}_n + L^{\M\pm}_n + L^{{\rm gh}\pm}_n .
\end{eqnarray*}
It satisfies the Virasoro algebra
\begin{eqnarray*}
   \left[ L_n^\pm, L_m^\pm \right] = (n-m)L_{n+m}^\pm  + \fr{c}{12}(n^3-n) \dl_{n+m,0} 
\end{eqnarray*}
and $\left[ L_n^+, L_m^- \right] = 0$, where the last term of the above is the central extension of the $SO(2,2)$ conformal algebra existing in 2 dimensional case only. The central charge of 2D quantum gravity vanishes as
\begin{eqnarray*}
     c = 1+ 6b_\L +c_\M -26=0 ,
\end{eqnarray*}
where $c_\M$ and $-26$ come from the matter and  $bc$-ghost fields, respectively. The $\phi$ field has the central charge $1+6b_\L$, in which $1$ comes from that $\phi$ is a scalar-like quantum boson field, while $6b_\L$ comes from that the Liouville action is not conformally invariant classically due to the presence of the $\hR\phi$ term. Actually, the Virasoro algebra at the Poisson bracket level has the non-zero central charge $6b_\L$. Thus, when quantized gravity, the conformal symmetry that is broken in a curved space quantum field theory is restored.

Now, the diffeomorphism invariance is realized as the conformal invariance and it implies that all theories connected by the conformal transformation are gauge-equivalent. In this way, we can show the background-metric independence, or the reparametrization invariance in 2 dimensions.

Let us discuss physical states in 2D quantum gravity. For a while, we disregard the ghost sector, which is considered to be integrated out here. First, we introduce the conformally invariant vacuum defined by the state satisfying the condition $L^{\L\pm}_n|\Om \rangle =0$ for $n \geq -1$, which is given by
\begin{eqnarray}
     |\Om \rangle = e^{-b_\L \phi_0}|0 \rangle ,
         \label{Liouville conformal vacuum}
\end{eqnarray}
where $\phi_0={\hat q}/\sq{2b_\L}$ is the zero-mode and $|0\rangle$ is the usual Fock vacuum that vanishes for all annihilation modes. The exponential factor has the origin in the $\hR \phi$ term, whose exponent is called the Liouville charge. Especially, the Liouville charge $-b_\L$ that the vacuum (\ref{Liouville conformal vacuum}) has is called the background charge.

We further introduce the Fock vacuum state with the Liouville charge $\gm$ defined by $|\gm \rang = e^{\gm \phi_0}|\Om \rangle$, which satisfies $H^\L |\gm \rang = h_\gm |\gm \rang$ with the eigenvalue $h_\gm = \gm-\gm^2/2b_\L$. We consider the state obtained by applying creation operators to this vacuum, such as
\begin{eqnarray*}
       | \Psi \rangle = {\cal O}(\a^{\pm\dag}_n,\cdots) |\gm \rang ,
\end{eqnarray*}
where the dots denote matter-field operators. The physical state is then defined by the conditions
\begin{eqnarray}
       \left( H^\L +  H^\M - 2 \right) |\Psi \rangle = 0, \quad 
       \left( L^{\L \pm}_n + L^{\M \pm}_n \right) |\Psi \rangle = 0  \,\,\,\, (n \geq 1) .
           \label{Liouville physical state condition}
\end{eqnarray}
This is nothing but the quantum version of the Wheeler-DeWitt equation. Here, note that $-2$ appears in the Hamiltonian condition, which reflects the spacetime dimensions in order to require that the spacetime volume integral of the corresponding field operator becomes diffeomorphism invariant.

For simplicity, we here consider physical states that primary fields in matter CFT are dressed by gravity. The real primary field with the same conformal weight $d$ for the left and right modes is defined by the conditions $L^{\M\pm}_0|d \rangle = d |d \rangle$ and $L^{\M\pm}_n|d \rangle =0$ for $n \geq 1$. Here, we denote this state using the primary conformal field $\Phi_d$ simply as $|d \rangle = \Phi_d^\dag |0\rangle$.\footnote{ 
Precisely, it is defined by the limit $|d \rangle = \lim_{\eta \to i\infty} e^{-i2d\eta}\Phi_d(\eta,\s) |0\rangle$.} 
The gravitationally dressed state is then given by
\begin{eqnarray*}
        \Phi_d^\dag |\gm_d \rangle  = e^{\gm_d \phi_0} \Phi^\dag_d |\Om \rang. 
\end{eqnarray*}
The Liouville charge $\gm_d$ is determined by solving the equation $h_{\gm_d} +2d = 2$ derived from the Hamiltonian condition in (\ref{Liouville physical state condition}). Here, the solution that approaches the classical value $2-2d$ at $b_\L \arr \infty$ is chosen as a physical gravitational value, which is given by 
\begin{eqnarray}
       \gm_d = b_\L \left( 1-\sq{ 1-\fr{4-4d}{b_\L} } \right) .
         \label{physical Liouville charge}
\end{eqnarray}

This state corresponds to the field operator $\l: \!e^{\gm_d \phi}\! \r: \! \Phi_d$, especially the $d=0$ operator $\l: \! e^{\gm_0 \phi} \! \r:$ gives the cosmological constant operator corresponding to $\sq{-g}$. The relationship between the state and the operator is given by the limit as follows: $\Phi_d^\dag |\gm_d \rangle = \lim_{\eta \arr i\infty}e^{-2i\eta} \l: \! e^{\gm_d \phi} \! \r: \! \Phi_d(\eta,\s) |\Om \rangle$.

From the duality relation $h_\gm = h_{2b_\L-\gm}$, we can find that the another solution state is given by $\Phi_d^\dag |2b_\L - \gm_d \rang$, which is called the dual state. This state itself has no physical meanings because there is no classical counterpart to it. Using the dual state, however, we can always introduce the norm structure such as $\lang 2b_\L-\gm_d |\gm_d \rang=1$ because the Liouville charges, including the total background charge $-2b_\L$, cancel out.

The correlation function among only the physical fields with (\ref{physical Liouville charge}) is defined by considering the interaction theory with the potential term such as the cosmological constant operator. The cancellation of the Liouville charge can be then achieved by the insertion of an appropriate fractional power of the potential operator in the correlation function. Such a correlation function can be evaluated using the method of analytic continuation \cite{gl,teschner}.

\subsection{BRST Operator and Physical States}

The $bc$-ghost action in the conformal gauge is given by $I_{\rm gh} = (-i/2\pi) \int d^2 x \sq{-\hg} b_{\mu\nu} \dl_c h^{\mu\nu} = (-i/\pi) \int d^2 x \sq{-\hg} b_{\mu\nu} \hnb^\mu c^\nu$, where $c^\mu$ is the ghost field and $b_{\mu\nu}$ is the traceless-symmetric antighost field.

We here introduce new field-variables $b_{\pm\pm} = b_{00} \pm b_{01}$ and $c^\pm = c^0 \pm c^1$ and rewrite the $bc$-ghost action in the form
\begin{eqnarray*}
   I_{\rm gh} = \fr{i}{\pi} \int  d^2 x \left( b_{++} \pd_- c^+ + b_{--} \pd_+ c^- \right) ,
\end{eqnarray*}
where $\pd_\pm = ( \pd_\eta \pm  \pd_\s )/2$. The equations of motion are then given by the simple forms $\pd_- c^+ = \pd_+ c^- =0$ and $\pd_- b_{++} = \pd_+ b_{--} =0$, and therefore we can mode-expand the fields as $c^\pm = \sum_{n \in {\bf Z}} c_n^\pm e^{-in (\eta \pm \s)}$ and $b_{\pm\pm} = \sum_{n \in {\bf Z}} b_n^\pm e^{-in (\eta \pm \s)}$. The Hermitian conjugates of these modes are defined by $c^{\pm\dag}_n=c^\pm_{-n}$ and $b^{\pm\dag}_n=b^\pm_{-n}$. Since the fields obey the equal-time anticommutation relations $\{ c^\pm(\s), b_{\pm\pm}(\s^\pp) \}= 2\pi \dl(\s-\s^\pp)$ and $\{ c^\pm(\s), b_{\mp\mp}(\s^\pp) \}=0$, the modes satisfy
\begin{eqnarray*}
     \{ c^\pm_n, b^\pm_m \} = \dl_{n+m,0},  \qquad \{c_n^\pm, b_m^\mp \} = 0 .
\end{eqnarray*}

The Virasoro generator for the ghost field is given by
\begin{eqnarray*}
   L_n^{{\rm gh}\pm} = \sum_{m \in {\bf Z}} (n + m) \l: b^\pm_{n-m} c^\pm_m \r: ,
\end{eqnarray*}
where we choose the normal ordering defined for the conformally invariant ghost vacuum $|0 \rang_{\rm gh}$ satisfying the conditions $L^{{\rm gh}\pm}_n |0 \rang_{\rm gh} =0 ~(n \geq -1)$, namely the annihilation modes setting as $c^\pm_n |0 \rang_{\rm gh}=0~(n \geq 2)$ and $b^\pm_n |0 \rang_{\rm gh}=0~(n \geq -1)$ are placed to the right.\footnote{ 
This is called the conformal normal ordering. On the other hand, what is called the creation-annihilation normal ordering is defined for the Fock ghost vacuum such that $c^\pm_n$ and $b^\pm_n$ with $n >0$ are placed to the right. If we write it by $\ddag ~\ddag$, the generator is described as $L^{{\rm gh}\pm}_n = \sum_{m \in {\bf Z}} (n+m) \ddag b^\pm_{n-m} c^\pm_m \ddag - \dl_{n,0}$. }  

The BRST operator imposing diffeomorphism invariance is given by $Q_{\rm BRST}= Q^+ + Q^-$, where
\begin{eqnarray*}
   Q^\pm = \sum_{n \in {\bf Z}} c_{-n}^\pm \left( L^{{\rm L} \pm}_n +  L^{{\rm M} \pm}_n \right) 
           - \half  \sum_{n,m \in {\bf Z}}  (n - m) \l: c_{-n}^\pm c_{-m}^\pm  b_{n+m}^\pm \r: .
\end{eqnarray*}
It can be decomposed as
\begin{eqnarray*}
   Q^\pm =c_0^\pm L_0^\pm - b_0^\pm M^\pm + \hat{Q}^\pm,  \qquad  M^\pm = 2 \sum_{n =1}^\infty  n c_{-n}^\pm c_n^\pm ,
\end{eqnarray*}
where
\begin{eqnarray*}
        \hat{Q}^\pm = \sum_{n \neq 0} c_{-n}^\pm \left( L_n^{{\rm L}\pm} + L_n^{{\rm M}\pm} \right)
                - \half  \sum_{n,m \neq 0 \atop n+m \neq 0}  (n - m) \l: c_{-n}^\pm c_{-m}^\pm  b_{n+m}^\pm \r: .
\end{eqnarray*}
The nilpotency of the BRST operator is then expressed as $\hat{Q}^{\pm 2} = L^\pm_0 M^\pm$'Æ$[ \hat{Q}^\pm, L^\pm_0 ]= [\hat{Q}^\pm, M^\pm] = [L^\pm_0, M^\pm]= 0$. The full Virasoro generator becomes BRST trivial such that $L_n^\pm = \{ Q_{\rm BRST}, b_n^\pm \}$.

In order to construct physical states, we introduce the Fock ghost vacuum $c^+_1 c^-_1 |0\rang_{\rm gh}$ annihilated by $c^\pm_n$ and $b^\pm_n$ with $n >0$.\footnote{ 
The state-operator correspondence for the Fock ghost vacuum is given by $c^+_1 c^-_1 |0 \rang_{\rm gh} = \lim_{\eta \arr i\infty}e^{2i\eta}c^+ c^-|0\rangle_{\rm gh}$.} 
We here consider the state obtained by applying the creation modes $\a^{\pm\dag}_n$, $c^{\pm\dag}_n$ and $b^{\pm\dag}_n$ with $n >0$ to the full Fock vacuum with the Liouville charge $\gm$ as follows:
\begin{eqnarray*}
      |\Psi \rang = {\cal O}(\a^{\pm\dag}_n, c^{\pm\dag}_n, b^{\pm\dag}_n, \cdots)  
                    |\gm \rang \otimes c^+_1 c^-_1 |0\rang_{\rm gh} .
\end{eqnarray*}
Physical states are then obtained by imposing the BRST invariance condition $Q_{\rm BRST} |\Psi \rang = 0$.

Since the state $|\Psi \rang$ annihilates when the zero-mode $b^\pm_0$ is applied to it, we only consider the subspace satisfying the conditions
\begin{eqnarray}
    b^\pm_0 |\Psi \rang =0, \qquad  L^\pm_0 |\Psi \rang = 0 ,
        \label{physical subspace in 2DQG}
\end{eqnarray}
where the second condition comes from that the full Hamiltonian is BRST-trivial such as $L^\pm_0=\{ Q_{\rm BRST}, b^\pm_0 \}$. Therefore, the BRST invariance condition on the subspace (\ref{physical subspace in 2DQG}) reduces to
\begin{eqnarray}
    \hat{Q}^\pm |\Psi \rang=0 .
      \label{Qhat condition}
\end{eqnarray}

As discussed before, if we consider the cases that ${\cal O}$ does not include the ghost mode, the condition (\ref{Qhat condition}) reduces to the Virasoro condition (\ref{Liouville physical state condition}). Here, the energy shift $-2$ arises from the algebra $L^{{\rm gh}\pm}_0 c^\pm_1 |0\rang_{\rm gh}= - c^\pm_1 |0\rang_{\rm gh}$.

In 2D quantum gravity, there are various physical states not discussed here, called the discrete states \cite{lz,bmp} and the grand ring state \cite{witten}. Furthermore, combining these states we can construct the $W_\infty$ current \cite{kp,klebanov}, and using this current we can derive the non-linear structure among correlation functions called the W-algebra constraint \cite{hamada94}.

Finally, we mention the norm of the ghost vacuum. Since the Hamiltonian $H^{\rm gh}=L^{{\rm gh}+}_0+L^{{\rm gh}-}_0$ is independent of the $c^\pm_0$ and $b^\pm_0$ modes, the ghost vacuum is degenerate such that ${}_{\rm gh}\lang 0 | 0 \rang_{\rm gh}=0$ as well as ${}_{\rm gh} \lang 0| c^-_{-1} c^+_{-1} c^+_1 c^-_1 |0 \rang_{\rm gh} = 0$, which can be easily shown by inserting $\{ b^\pm_0, c^\pm_0 \}=1$ into each vacuum norm. Therefore, using the degenerate pair given by applying $\vartheta = i c^+_0 c^-_0$ to the Fock ghost vacuum, the norm structure is introduced as ${}_{\rm gh} \lang 0| c^-_{-1} c^+_{-1} \vartheta c^+_1 c^-_1 |0 \rang_{\rm gh}=1$.

\section{4D Quantum Gravity on $R \times S^3$}
\setcounter{equation}{0}

We here review recent developments on the BRST conformal symmetry for 4D quantum gravity on the cylindrical background $R \times S^3$ \cite{hamada12b}.

Unlike 2 dimensional case, the traceless tensor field $h_{\mu\nu}$ has dynamical degrees of freedom. Its dynamics is described by the conformally invariant Weyl action divided by $t^2$ as 
\begin{eqnarray*}
   I_{\rm W} = -\fr{1}{t^2} \int d^4 x \sq{-g} C_{\mu\nu\lam\s}^2 , 
\end{eqnarray*}
where the coupling constant $t$ is introduced in the expansion of the metric (\ref{metric decomposition}). At the $t \to 0$ limit, the kinetic term of the traceless tensor field remains. In the following, we consider this limit.

The dynamics of the $\phi$ field is described by the Wess-Zumino action induced from the path integral measure as in the case of 2D quantum gravity. At $t \to 0$ limit, it is given by the Riegert action \cite{riegert} defined by
\begin{eqnarray}
   S_\R &=& - \fr{b_c}{(4\pi)^2} \int d^4 x \int^\phi_0 d\phi \sq{-g} E_4   
               \nonumber \\
        &=& - \fr{b_c}{(4\pi)^2} \int d^4 x \sq{-\hg} \left( 2 \phi \hDelta_4 \phi + \hE_4 \phi \right) ,
           \label{Riegert action}
\end{eqnarray}
where $\sq{-g}\Delta_4$ is the conformally invariant 4th-order differential operator for a scalar given by
\begin{eqnarray*}
    \Delta_4 = \nb^4 +2 R^{\mu\nu} \nb_\mu \nb_\nu - \fr{2}{3} R \nb^2 + \fr{1}{3} \nb^\mu R \nb_\mu .
\end{eqnarray*}
The quantity $E_4$ is the extended Euler density defined by\footnote{ 
From the renormalization group analysis of quantum field theories in curved space \cite{hathrellS, hathrellQED, freeman}, we can show that this combination indeed appears as a conformal anomaly at all orders \cite{hamada14a, hm16}.} 
\begin{eqnarray*}
   E_4= G_4- \fr{2}{3} \nb^2 R , 
\end{eqnarray*}
where $G_4=R^2_{\mu\nu\lam\s}-4R^2_{\mu\nu}+R^2$ is the usual Euler density. It is significant that this combination satisfies $\sq{-g}E_4 = \sq{-\hg}(4 \hDelta_4 \phi + \hE_4)$ for $g_{\mu\nu}=e^{2\phi}\hg_{\mu\nu}$ as similar to the corresponding equation for $\sq{-g}R$ in 2 dimensions. The coefficient in front of the Riegert action is determined to be 
\begin{eqnarray}   
   b_c= \fr{1}{360} \left( N_S+11N_F+62N_A \right) + \fr{769}{180} ,
     \label{coefficient of Riegert action}
\end{eqnarray} 
where the first three terms come from $N_S$ scalars, $N_F$ fermions and $N_A$ gauge fields \cite{duff} and the last term comes from the quantized gravitational fields \cite{ft,amm92}.

The action of 4D quantum gravity is given by the sum of the Riegert action and the action $I$ composed of the Weyl action and the matter field actions. The kinetic term of the Weyl action has the invariance under the gauge transformation $\dl_\kappa h_{\mu\nu} = \hnb_\mu \kappa_\nu + \hnb_\nu \kappa_\mu - \hg_{\mu\nu} \hnb_\lam \kappa^\lam/2$, where $\kappa^\mu=\xi^\mu/t$. We will fix these gauge degrees of freedom completely. Even after the gauge-fixing, the finite degrees of freedom $\kappa^\mu=\zeta^\mu$ satisfying the conformal Killing equation $\hnb_\mu \zeta_\nu + \hnb_\nu \zeta_\mu - \hg_{\mu\nu} \hnb_\lam \zeta^\lam/2 =0$ still survive.  These residual gauge degrees of freedom generate the conformal transformations (\ref{conformal transformations with zeta^mu}).

\subsection{Canonical Quantizations of Gravitational Fields}

Let us quantize 4D quantum gravity action $S_{\rm 4DQG}=S_{\rm R}+I_{\rm W}+I_{\rm M}$ on $R \times S^3$ \cite{amm97b,hh,hamada09}, where $I_{\rm M}$ is a conformally coupled matter field action, which is not specified here and below. The background metric $\hg_{\mu\nu}$ is parametrized by the coordinate $x^\mu=(\eta,x^k)$ using the Euler angles $x^k=(\a,\b,\gm)$. The spatial metric is then defined by $\hgm_{kl} dx^k dx^l = (1/4)(d\a^2 +d\b^2 +d\gm^2 +2 \cos \b d\a d\gm )$, where $\a$, $\b$ and $\gm$ have the ranges $[0,2\pi]$, $[0,\pi]$, and $[0,4\pi]$, respectively. The radius of $S^3$ is taken to be unity such that $\hR=6$. The volume element on the unit $S^3$ is $d\Om_3 = \sin \b d\a d\b d\gm/8$ and the volume is given by $\V3 = \int d\Om_3 =2\pi^2$.

First, we decompose the traceless tensor field as $h_{00}$, $h_{0k}$ and $h_{kl} = h^{\bf tr}_{kl} + \hgm_{kl} h_{00}/3$, where $ h^{\bf tr}_{kl}$ denotes spatial traceless components. We here take the radiation gauge defined by 
\begin{eqnarray*}
    h_{00}=\hnb^k h_{0k} =\hnb^k h^{{\bf tr}}_{kl}= 0 .
\end{eqnarray*}
Introducing new variables $\h_{kl}$ for the transverse-traceless tensor and $\h_k$ for the transverse vector components, 
the gauge-fixed action is then given by
\begin{eqnarray*}
    S_{\rm 4DQG}   &=&  \int d\eta  \int_{S^3}  d\Om_3 \biggl\{ 
        -\fr{2b_c}{(4\pi)^2}\phi \left( 
        \pd_\eta^4 -2\Box_3 \pd_\eta^2 +\Box_3^2 + 4\pd_\eta^2 \right) \phi 
                 \nonumber  \\ 
    &&  -\fr{1}{2} \h_{kl} \left( \pd_\eta^4 -2\Box_3 \pd_\eta^2 + \Box_3^2 
        + 8\pd_\eta^2-4\Box_3 +4 \right) \h^{kl} 
                 \nonumber    \\
    &&  + \h_k \left( \Box_3 +2 \right) 
        \left( -\pd_\eta^2 +\Box_3 -2 \right) \h^k   \biggr\} ,
\end{eqnarray*}
where $\Box_3 =\hgm^{kl}\hnb_k \hnb_l$ is the Laplacian on $S^3$.

Furthermore, we remove the mode in the transverse vector field satisfying $(\Box_3 + 2)\h_k=0$, because the kinetic term for this mode vanishes. The radiation gauge with this extra condition is called the radiation$^+$ gauge, which is represented as \cite{hh}
\begin{eqnarray}
   \h_k |_{J=\half}=0
      \label{radiation plus condition}
\end{eqnarray}
using the mode as will be introduced in (\ref{mode expansions of tensor and vector fields}) below.

The gravitational fields are mode-expanded using the symmetric-transverse-traceless (ST$^2$) $n$-th rank spherical tensor harmonics on $S^3$ \cite{ro}, denoted by $Y^{k_1 \cdots k_n}_{J(M \veps_n)}$ satisfying the equation $\Box_3 Y^{k_1 \cdots k_n}_{J(M \veps_n)} = [-2J(2J+2)+n]Y^{k_1 \cdots k_n}_{J(M \veps_n)}$ \cite{hh}. It belongs to the $(J+\veps_n,J-\veps_n)$ representation of the $S^3$ isometry group $SU(2) \times SU(2)$ with $J \geq n/2$ and the multiplicity $M=(m,m^\pp)$ with the polarization index $\veps_n = \pm n/2$, where $m=-J-\veps_n, \cdots , J+\veps_n$ and $m^\pp = -J+\veps_n, \cdots , J-\veps_n$ and thus the multiplicities are given by $2(2J+n+1)(2J-n+1)$ for $n >0$ and $(2J+1)^2$ for $n=0$. The complex conjugate is defined by $Y^{k_1 \cdots k_n \,*}_{J(M \veps_n)} = (-1)^n \eps_M Y^{k_1 \cdots k_n}_{J(-M \veps_n)}$, where $\eps_M =(-1)^{m-m\pp}$, and the normalization is taken as $\int d\Om_3 Y^{k_1 \cdots k_n \,*}_{J_1(M_1 \veps_n^1)} Y_{k_1 \cdots k_n \,J_2 (M_2 \veps_n^2)} = \dl_{J_1 J_2} \dl_{M_1 M_2}\dl_{\veps_n^1 \veps_n^2}$, where $\dl_{M_1 M_2}= \dl_{m_1 m_2}\dl_{m_1^\pp m_2^\pp}$. In the following, we use the notations $y=\veps_1=\pm 1/2$ and $x=\veps_2 = \pm 1$ for the polarization indices of vector and rank 2 tensor.

From the equations of motion on $R \times S^3$, we find that the $\phi$ field is expanded using the scalar harmonics on $S^3$ as
\begin{eqnarray*}
   \phi  &=& \fr{\pi}{2\sq{b_c}} \biggl\{ 2({\hat q} + {\hat p}\eta ) Y_{00} 
                 \nonumber \\
       &&  + \sum_{J \geq \half} \sum_M   \!  \fr{1}{\sq{J(2J + 1)}} 
           \left( a_{JM}e^{-i2J\eta} Y_{JM}  + a^\dag_{JM} e^{i2J\eta}Y_{JM}^*  \right)
                \nonumber \\
       &&  + \sum_{J \geq 0} \sum_M   \!  \fr{1}{\sq{(J + 1)(2J + 1)}} 
           \left(  b_{JM}e^{-i(2J+2)\eta} Y_{JM}  + b^\dag_{JM} e^{i(2J+2)\eta}Y_{JM}^*  \right)  
           \biggr\} .
\end{eqnarray*}
The quantization can be carried out using the standard Dirac's method for higher-derivative fields. We then obtain the commutation relations for the mode operators as
\begin{eqnarray*}
    \bigl[ {\hat q}, {\hat p} \bigr] = i,   \quad 
    \bigl[ a_{J_1 M_1},a^\dag_{J_2 M_2} \bigr] = - \bigl[ b_{J_1 M_1},b^\dag_{J_2 M_2} \bigr]  
    = \dl_{J_1 J_2}\dl_{M_1 M_2} .
\end{eqnarray*}
Therefore, $a_{JM}$ and $b_{JM}$ are the positive-metric and negative-metric modes, respectively. The Hamiltonian is given by
\begin{eqnarray*}
   H = \half {\hat p}^2 + b_c + \sum_{J \geq 0} \sum_M 
       \{ 2J a^\dag_{JM} a_{JM} -(2J+2)b^\dag_{JM} b_{JM} \} .
\end{eqnarray*}
Here, the constant energy shift $b_c$ is the Casimir effect on $R \times S^3$, which is here determined as the conformal algebra (\ref{conformal algebra on RxS^3}) given below closes quantum mechanically.

Similarly, the transverse-traceless tensor and the transverse vector fields are expanded using the ST$^2$ tensor and vector harmonics as\footnote{ 
The use of this unusual convention for $\h^k$ is to match with the convention of conformal algebra for tensor modes given later \cite{hamada09}.}
\begin{eqnarray}
   \h^{kl} &=&  \fr{1}{4} \sum_{J \geq 1}\sum_{M,x}   \fr{1}{\sq{J(2J + 1)}} \Bigl\{
       c_{J(Mx)} e^{-i2J\eta}Y^{kl}_{J(Mx)}
       + c^{\dag}_{J(Mx)} e^{i2J\eta}Y^{kl*}_{J(Mx)} \Bigr\}
              \nonumber   \\
   &&  + \fr{1}{4} \sum_{J \geq 1} \sum_{M,x}  \fr{1}{\sq{(J + 1)(2J + 1)}} \Bigl\{
       d_{J(Mx)} e^{-i(2J + 2)\eta} Y^{kl}_{J(Mx)}
                \nonumber   \\
   &&  \qquad 
       + d^{\dag}_{J(Mx)} e^{i(2J + 2)\eta}Y^{kl*}_{J(Mx)} \Bigr\},
                \nonumber  \\
     \h^k  &=& i \fr{1}{2}\sum_{J \geq 1}\sum_{M,y}  \fr{1}{\sq{(2J - 1)(2J + 1)(2J + 3)}}
                \nonumber \\
   &&  \times  
       \Bigl\{   e_{J(My)} e^{-i(2J + 1)\eta}Y^k_{J(My)}
        -  e^{\dag}_{J(My)} e^{i(2J + 1)\eta}Y^{k*}_{J(My)}  \Bigr\} ,
           \label{mode expansions of tensor and vector fields}
\end{eqnarray}
where $\h_k$ is expanded by the modes with $J \geq 1$ reflecting the radiation$^+$ gauge condition (\ref{radiation plus condition}). The commutation relations are given by
\begin{eqnarray*}
    \bigl[ c_{J_1 (M_1 x_1)}, c^\dag_{J_2 (M_2 x_2)}  \bigr]  &=&  - \bigl[ d_{J_1 (M_1 x_1)}, d^\dag_{J_2 (M_2 x_2)} \bigr]
    =  \dl_{J_1 J_2} \dl_{M_1 M_2}\dl_{x_1 x_2},
             \nonumber   \\
    \bigl[  e_{J_1 (M_1 y_1)}, e^\dag_{J_2 (M_2 y_2)} \bigr]   &=&  -\dl_{J_1 J_2}\dl_{M_1 M_2}\dl_{y_1 y_2} .
\end{eqnarray*}
Thus, $c_{JM}$ has the positive-metric, while $d_{JM}$ and $e_{JM}$ have the negative-metric. The Hamiltonian is given by
\begin{eqnarray*}
   H  &=&  \sum_{J \geq 1}\sum_{M,x} \{ 2J c^{\dag}_{J(Mx)}c_{J(Mx)}
          -(2J+2)d^{\dag}_{J(Mx)}d_{J(Mx)} \}
                \nonumber  \\
   &&    -\sum_{J \geq 1}\sum_{M,y} (2J+1) e^{\dag}_{J(My)}e_{J(My)} .
\end{eqnarray*}

\subsection{Generators of Conformal Symmetry}

The 15 conformal Killing vectors on $R \times S^3$ \cite{amm97a} satisfy the equations $3 \pd_\eta \zeta_0 + \psi = 0$, $\pd_\eta \zeta_k + \hnb_k \zeta_0 = 0$ and $\hnb_k \zeta_l + \hnb_l \zeta_k - 2 \hgm_{kl} \psi/3 = 0$, where $\psi= \hnb_k \zeta^k$. The solutions satisfying $\psi=0$, namely the Killing vectors are given by $\zeta^\mu_\T=(1,0,0,0)$ and $\zeta^\mu_\R=(0, \zeta^k_\R)$ with

\begin{eqnarray*}
    ( \zeta^k_\R )_{MN} 
      = i \fr{\V3}{4} \left\{ Y^*_{\half M} \hnb^k Y_{\half N} 
              - Y_{\half N} \hnb^k Y^*_{\half M}  \right\} .
\end{eqnarray*}
The solutions satisfying $\psi \neq 0$ are given by $\zeta^\mu_\S=(\zeta^0_\S, \zeta^k_\S)$ with
\begin{eqnarray*} 
   (\zeta_\S^0)_M = \fr{\sq{\V3}}{2} \e^{i\eta}Y^*_{\half M} , \quad 
   (\zeta_\S^k)_M = -i \fr{\sq{\V3}}{2} \e^{i\eta} \hnb^k Y^*_{\half M} 
\end{eqnarray*}
and its complex conjugate $\zeta^{\mu *}_\S$. Here and below, for simplicity, we use the indices $M$, $N$ without $J$ for the ${\bf 4}$-vector of $J=1/2$ which appears in the conformal Killing vectors, the corresponding generators, and also ghost modes introduced below.

The generator of diffeomorphism symmetry that forms the conformal algebra is given by $Q_\zeta = \int d\Om_3 \zeta^\mu \l: \hat{T}_{\mu 0} \r:$, where the energy-momentum tensor is defined by ${\hat T}^{\mu\nu}= (2/\sq{-\hg})\dl S_{\rm 4DQG}/\dl \hg_{\mu\nu}$. In the following, the generators for the conformal Killing vectors $\zeta^\mu= \zeta_\T^\mu, (\zeta_\R^\mu)_{MN}, (\zeta_\S^\mu)_M, (\zeta_\S^{\mu*})_M$ are denoted by $H$, $R_{MN}$, $Q_M$ and $Q^\dag_M$, respectively. $H$ is the Hamiltonian given before, which corresponds to the dilatation generator on $R \times S^3$. $R_{MN}$ are the 6 rotation generators on $S^3$ satisfying the relations $R^\dag_{MN}=R_{NM}$ and $R_{MN}=-\eps_M \eps_N R_{-N-M}$, whose explicit forms are not depicted here. $Q_M$ are the 4 generators of the special conformal transformations and their Hermite conjugates $Q_M^\dag$ corresponds to the 4 generators of translation on $R \times S^3$.

According to the definition above, we can construct the generator of the special conformal transformation for the Riegert sector as \cite{amm97b}
\begin{eqnarray}
     Q_M  &=& \left( \hbox{$\sq{2b_c}$}-i\hat{p} \right) a_{\half M}
           +\sum_{J \geq 0}  \sum_{M_1}  \sum_{M_2} \C^{\half M}_{JM_1, J+\half M_2}
               \nonumber \\
     &&  \times
           \Bigl\{ \a(J) \eps_{M_1} a^\dag_{J-M_1} a_{J+\half M_2}
           +\b(J) \eps_{M_1} b^\dag_{J-M_1} b_{J+\half M_2}
               \nonumber  \\
     &&  \quad\
           + \eps_{M_2} a^\dag_{J+\half -M_2} b_{J M_1} \Bigr\} ,
         \label{Q_M for Riegert sector}
\end{eqnarray}
where
\begin{eqnarray}
    \a(J) =  \sq{2J(2J+2)},  \qquad   \b(J) = -\sq{(2J+1)(2J+3)} .
       \label{alpha and beta coefficients of Q_M}
\end{eqnarray}
The $\C$-coefficient is the $SU(2) \times SU(2)$ Clebsch-Gordan coefficient defined by
\begin{eqnarray*}
   \C^{JM}_{J_1M_1,J_2M_2} &=& \sq{\V3}  \int_{S^3} d\Om_3 Y^*_{JM}Y_{J_1M_1}Y_{J_2M_2}
                 \nonumber \\
          &=& \sq{\fr{(2J_1+1)(2J_2+1)}{2J+1}} C^{Jm}_{J_1m_1,J_2m_2} C^{J\prm}_{J_1\prm_1,J_2\prm_2} ,
\end{eqnarray*} 
where $C^{Jm}_{J_1m_1,J_2m_2}$ is the standard Clebsch-Gordan coefficient \cite{vmk}, and thus $J+J_1+J_2$ is integer and the triangle inequality $|J_1 -J_2| \leq J \leq J_1 +J_2$ and $M=M_1 +M_2$ are satisfied. It is a real function satisfying the relations $\C^{JM}_{J_1M_1,J_2M_2}=\C^{JM}_{J_2M_2,J_1M_1}=\C^{J-M}_{J_1-M_1,J_2-M_2}=\eps_{M_2}\C^{J_1 M_1}_{JM,J_2 -M_2}$, $\C^{00}_{JM,JN}= \eps_M \dl_{M-N}$, and the crossing relation 
\begin{eqnarray}
    \sum_{J \geq 0} \sum_M \eps_M \C^{J_1M_1}_{J_2M_2, J-M}  \C^{J_3M_3}_{JM,J_4M_4} 
    = \sum_{J \geq 0} \sum_M \eps_M \C^{J_1M_1}_{J_4M_4, J-M} \C^{J_3M_3}_{JM,J_2M_2}  . 
            \label{crossing relations}                        
\end{eqnarray}
For $Q_M$, the $\C$-coefficient with $J=1/2$ appears.

For the traceless tensor field, we obtain \cite{hh}
\begin{eqnarray}
    Q_M  &=&  \sum_{J \geq 1} \sum_{M_1,x_1} \sum_{M_2,x_2} \E^{\half M}_{J(M_1 x_1), J + \half (M_2 x_2)}
           \Bigl\{ \a(J) \eps_{M_1} c^\dag_{J(-M_1 x_1)} c_{J + \half (M_2 x_2)}
                        \nonumber  \\
    &&  \quad
          + \b(J) \eps_{M_1} d^\dag_{J(-M_1 x_1)} d_{J + \half (M_2 x_2)}
          + \eps_{M_2} c^\dag_{J + \half (-M_2 x_2)} d_{J (M_1 x_1)} \Bigr\}
                        \nonumber \\
    &&    + \! \sum_{J \geq 1} \sum_{M_1,x_1} \sum_{M_2,y_2} \H^{\half M}_{J(M_1 x_1); J (M_2 y_2)}
                        \nonumber  \\
    && \quad \times
          \Bigl\{ A(J) \eps_{M_1} c^\dag_{J(-M_1 x_1)} e_{J (M_2 y_2)}
          +  B(J) \eps_{M_2} e^\dag_{J(-M_2 y_2)} d_{J (M_1 x_1)}   \Bigr\}
                     \nonumber  \\
    &&    +  \sum_{J \geq 1}  \sum_{M_1,y_1}  \sum_{M_2, y_2}
          \D^{\half M}_{J(M_1 y_1), J + \half (M_2 y_2)}
          C(J) \eps_{M_1} e^\dag_{J(-M_1 y_1)} e_{J + \half (M_2 y_2)} ,
       \label{Q_M for Weyl sector}
\end{eqnarray}
where $\a(J)$ and $\b(J)$ are the same as (\ref{alpha and beta coefficients of Q_M}). The other coefficients are given by
\begin{eqnarray*}
     A(J) &=&  \sq{\fr{4J}{(2J-1)(2J+3)}}, \qquad  B(J) = \sq{\fr{2(2J+2)}{(2J-1)(2J+3)}},
                  \nonumber   \\
     C(J) &=&  \sq{\fr{(2J-1)(2J+1)(2J+2)(2J+4)}{2J(2J+3)}} .
\end{eqnarray*}
The $SU(2) \times SU(2)$ Clebsch-Gordan coefficients with tensor indices are defined by 
\begin{eqnarray*}
   \E^{J M}_{J_1(M_1 x_1), J_2(M_2 x_2)} &=& \sq{\V3}  \int_{S^3} d\Om_3 Y^*_{J M}Y^{kl}_{J_1(M_1x_1)}Y_{kl J_2(M_2x_2)},
         \nonumber \\
   \H^{J M}_{J_1(M_1 x_1); J_2(M_2 y_2)} &=& \sq{\V3} \int_{S^3} d\Om_3 Y^*_{J M}Y^{kl}_{J_1(M_1x_1)}\hnb_k Y_{l J_2(M_2y_2)},
         \nonumber \\
   \D^{J M}_{J_1(M_1 y_1), J_2(M_2 y_2)} &=& \sq{\V3}  \int_{S^3} d\Om_3 Y^*_{JM}Y^k_{J_1(M_1y_1)}Y_{k J_2(M_2y_2)} .
\end{eqnarray*}
The general forms of these coefficients can be represented using 6-j symbols, which are calculated in \cite{hh}. For the special cases that appear in the generator, they are given by
\begin{eqnarray*}
     \E^{\half M}_{J(M_1 x_1), J+\half (M_2 x_2)}
       &=& \sq{(2J - 1)(J + 2)}  C^{\half m}_{J+x_1 m_1, J+\half+x_2 m_2} 
           C^{\half \prm}_{J-x_1 \prm_1, J+\half-x_2 \prm_2}  ,
                \nonumber \\ 
    \H^{\half M}_{J(M_1 x_1); J(M_2 y_2)} 
       &=& - \sq{(2J - 1)(2J + 3)}  C^{\half m}_{J+x_1 m_1, J+y_2 m_2} 
           C^{\half \prm}_{J-x_1 \prm_1, J-y_2 \prm_2} ,
              \nonumber \\
    \D^{\half M}_{J(M_1 y_1), J+\half (M_2 y_2)}
      &=& \sq{J(2J + 3)} C^{\half m}_{J+y_1 m_1, J+\half+y_2 m_2} 
           C^{\half \prm}_{J-y_1 \prm_1, J+\half-y_2 \prm_2} .
\end{eqnarray*}

The 15 generators form the conformal algebra of $SO(4,2)$ as follows:\footnote{ 
Parametrizing the ${\bf 4}$-vector index $\{ (1/2, 1/2), (1/2, -1/2), (-1/2, 1/2),  (-1/2, -1/2) \}$ by $\{ 1,2,3,4 \}$, and setting $A_+=R_{31}$, $A_-=R_{31}^\dag$, $A_3=(R_{11}+R_{22})/2$, $B_+=R_{21}$, $B_-=R_{21}^\dag$ and $B_3=(R_{11}-R_{22})/2$, the last rotation algebra is written in the familiar form of the $SU(2)\times SU(2)$ algebra as $[A_+, A_-]=2A_3$, $[A_3, A_\pm ]=\pm A_\pm$, $[B_+, B_-]=2B_3$, $[B_3, B_\pm ]=\pm B_\pm$, where $A_{\pm,3}$ and $B_{\pm,3}$ commute. The generators $A_{\pm,3} (B_{\pm,3})$ act on the left(right) index of $M=(m,m^\pp)$. 
} 
\begin{eqnarray}
     \bigl[ Q_M, Q^\dag_N \bigr]  &=&  2\dl_{MN} H + 2R_{MN},
           \nonumber \\
    \bigl[ H, Q_M \bigr]  &=& -Q_M,   \qquad   \bigl[ H, Q_M^\dag \bigr]  = Q_M^\dag,
           \nonumber \\
    \bigl[ H, R_{MN} \bigr]  &=&  \bigl[ Q_M, Q_N \bigr] = 0,
           \nonumber  \\
    \bigl[ Q_M, R_{M_1 M_2} \bigr]  &=&  \dl_{M M_2}Q_{M_1}  -\eps_{M_1}\eps_{M_2}\dl_{M -M_1}Q_{-M_2} ,
            \nonumber  \\
    \bigl[ R_{M_1 M_2}, R_{M_3 M_4} \bigr]  &=&  \dl_{M_1 M_4} R_{M_3 M_2} 
            -\eps_{M_1}\eps_{M_2} \dl_{-M_2 M_4} R_{M_3 -M_1}
            \nonumber \\
     && - \dl_{M_2 M_3} R_{M_1 M_4}  + \eps_{M_1}\eps_{M_2} \dl_{-M_1 M_3} R_{-M_2 M_4} .
        \label{conformal algebra on RxS^3}
\end{eqnarray}

The significant property of the $Q_M$ generators given by (\ref{Q_M for Riegert sector}) and (\ref{Q_M for Weyl sector}) is that these generators mix the positive-metric and negative-metric modes due to the presence of the cross terms. Consequently, the negative-metric mode cannot be gauge-invariant alone, and therefore it itself has no physical meanings.

\subsection{BRST Formalism}

The BRST conformal transformation is defined by replacing the gauge parameter $\zeta^\mu$ in (\ref{conformal transformations with zeta^mu}) with the corresponding gauge ghost $c^\mu$ satisfying the conformal Killing equation $\hnb_\mu c_\nu + \hnb_\nu c_\mu - \hg_{\mu\nu} \hnb_\lam c^\lam/2 =0$. The gauge ghost is here expanded by 15 Grassmann modes $\rc$, $\rc_{MN}$, $\rc_M$, $\rc_M^\dag$ as
\begin{eqnarray*}
     c^\mu = \rc \eta^\mu + \sum_M \left( \rc_M^\dag \zeta^\mu_M + \rc_M \zeta_M^{\mu *} \right)
              + \sum_{M,N} \rc_{MN} \zeta_{MN}^\mu .
\end{eqnarray*}
Here, $\rc$ is a real operator, and $\rc_{MN}$ satisfies the relations $\rc_{MN}^\dag = \rc_{NM}$ and $\rc_{MN} = -\eps_M \eps_N \rc_{-N-M}$. We also introduce the antighost modes $\rb$, $\rb_{MN}$, $\rb_M$, $\rb_M^\dag$ that have the same properties as the ghost modes. The anticommutation relations among these modes are defined by
\begin{eqnarray*}
     &&  \{ \rb, \rc \} = 1, \qquad   
         \{ \rb_{M N}, \rc_{L K} \} = \dl_{M L} \dl_{N K} 
                                      -\eps_M \eps_N \dl_{-M K} \dl_{-N L} ,
              \nonumber \\
     &&  \{ \rb^\dag_M , \rc_N \} = \{ \rb_M , \rc^\dag_N \} = \dl_{MN} .
\end{eqnarray*}

Using these gauge ghost and antighost modes, we can construct the 15 generators of conformal symmetry as follows \cite{amm97a}:
\begin{eqnarray}
    H^\gh  &=&  \sum_M \left( \rc_M^\dag \rb_M - \rc_M \rb_M^\dag \right) ,
                      \nonumber \\
    R^\gh_{MN}  &=&  - \rc_M \rb_N^\dag + \rc^\dag_N \rb_M 
                     + \eps_M \eps_N \left( \rc_{-N} \rb^\dag_{-M} - \rc^\dag_{-M} \rb_{-N} \right)
                       \nonumber \\
                &&   - \sum_L \left( \rc_{LM} \rb_{LN} - \rc_{NL} \rb_{ML} \right) ,
                       \nonumber \\
    Q_M^\gh &=&  -2 \rc_M \rb - \rc \rb_M - \sum_L \left( 2 \rc_{LM} \rb_L  + \rc_L \rb_{ML} \right),
                       \nonumber \\
    Q_M^{\gh\dag} &=& 2 \rc_M^\dag \rb + \rc \rb_M^\dag 
                      + \sum_L \left( 2 \rc_{ML} \rb_L^\dag + \rc_L^\dag \rb_{LM} \right) .
           \label{generators on RxS^3 for ghost sector}
\end{eqnarray}
These generators satisfy the same conformal algebra as (\ref{conformal algebra on RxS^3}). In the following, we write the full generators of the conformal algebra including the gauge ghost part as
\begin{eqnarray*}
    {\cal H}  &=&  H + H^\gh, \qquad  {\cal R}_{MN} = R_{MN} + R^\gh_{MN} ,
                \nonumber \\
    {\cal Q}_M  &=&   Q_M + Q_M^\gh, \qquad  {\cal Q}^\dag_M = Q^\dag_M + Q_M^{\gh\dag} .
\end{eqnarray*}

The nilpotent BRST operator generating the diffeomorphism (\ref{conformal transformations with zeta^mu}) is now given by \cite{hamada12b}
\begin{eqnarray*}
    &&  Q_\BRST = \rc H + \sum_M \left( \rc_M^\dag Q_M + \rc_M Q_M^\dag \right) + \sum_{M,N} \rc_{MN}R_{MN} 
                     \nonumber \\
    &&  \quad
          + \half \rc H^\gh + \half \sum_M \left( \rc^\dag_M Q_M^\gh 
          + \rc_M Q_M^{\gh \dag} \right) + \half \sum_{M,N} \rc_{MN} R_{MN}^\gh .
\end{eqnarray*}
It can be written in the form
\begin{eqnarray*}
   Q_\BRST = \rc {\cal H} + \sum_{M,N} \rc_{MN} {\cal R}_{MN}
               - \rb M - \sum_{M,N} \rb_{MN} Y_{MN} + \hat{Q}  ,
\end{eqnarray*}
where ${\cal H}$ and ${\cal R}_{MN}$ are the full generators and the other operators are defined by
\begin{eqnarray*}
      M &=& 2 \sum_M \rc^\dag_M \rc_M ,  \qquad  Y_{MN} = \rc^\dag_M \rc_N + \sum_L \rc_{ML} \rc_{LN} ,
                     \nonumber \\
     \hat{Q} &=& \sum_M \left( \rc_M^\dag Q_M + \rc_M Q_M^\dag \right)  .
\end{eqnarray*}
The nilpotency $Q_\BRST^2=0$ can be proved using the conformal algebra (\ref{conformal algebra on RxS^3}).

The anticommutation relations of the BRST operator with the antighost modes are calculated as
\begin{eqnarray}
   && \left\{ Q_\BRST, \rb \right\} = {\cal H} , \qquad  \left\{ Q_\BRST, \rb_{MN} \right\} = 2 {\cal R}_{MN},
                 \nonumber \\
   && \left\{ Q_\BRST, \rb_M \right\} = {\cal Q}_M , \qquad  \left\{ Q_\BRST, \rb^\dag_M \right\} = {\cal Q}_M^\dag ,
       \label{BRST triviallity of generator}
\end{eqnarray}
where the full generators appear in the right-hand side. This implies that the descendant states obtained by applying ${\cal Q}_M^\dag$ to the BRST invariant primary states discussed below become BRST trivial.

The BRST-invariant physical state $|\Psi \rangle$ is defined by the condition
\begin{eqnarray*}
     Q_\BRST |\Psi \rangle  = 0 .
\end{eqnarray*}
In the following, we analyze this condition and find an infinite number of physical states that correspond to various diffeomorphism invariant scalar combinations of curvature tensors.

In order to construct physical states, we introduce some vacuum states. The vacuum of Fock space annihilated by the zero mode $\hat{p}$ and annihilation modes $a_{JM}$ and $b_{JM}$ is denoted by $|0\rangle$. We also introduce the conformally invariant vacuum annihilated by all generators of the conformal symmetry except gauge ghost parts, $H$, $R_{MN}$, $Q_M$, and $Q_M^\dag$, which is defined by  
\begin{eqnarray*}
     | \Om \rang = e^{-2b_c \phi_0(0)} \rvac ,
\end{eqnarray*}
where $\phi_0(0)={\hat q}/\sq{2b_c}$. This vacuum and its Hermitian conjugate have the background charge $-2b_c$, respectively, and thus the total background charge of the vacua is $-4b_c$.\footnote{ 
The background charge originates from the linear term in the Riegert action (\ref{Riegert action}).
} 

The conformally invariant vacuum of the gauge ghost sector is denoted by $|0 \rangle_\gh$, which is annihilated by all generators of the gauge ghost system (\ref{generators on RxS^3 for ghost sector}); namely, annihilated by all antighosts, but not annihilated by gauge ghosts. The Fock ghost vacuum annihilated by the annihilation modes $\rc_M$ and $\rb_M$ is then defined by $\prod \rc_M |0 \rangle_\gh$.

Since the Hamiltonian depends on neither $\rc$ and $\rc_{MN}$ nor $\rb$ and $\rb_{MN}$, the gauge ghost vacuum $\prod \rc_M|0 \rangle_\gh$ is degenerate. The degenerate partners are then given by applying $\rc$ and $\prod \rc_{MN}$ to this vacuum. The norm structure of the vacua will be discussed later.

For convenience, we denote the Fock vacuum state with the Riegert charge $\gm$, including the ghost vacuum, by 
\begin{eqnarray}
       | \gm \rangle = e^{\gm \phi_0(0)} | \Om \rangle \otimes \prod_M \rc_M |0 \rangle_\gh .
            \label{state with gamma}
\end{eqnarray}
This state satisfies ${\cal H} |\gm \rangle = (h_\gm -4) |\gm \rangle$, where
\begin{eqnarray}
        h_\gm = \gm - \fr{\gm^2}{4b_c} 
        \label{definition of conformal weight h_a}
\end{eqnarray} 
and $-4$ comes from the gauge ghost sector.

Physical states are constructed by applying the creation modes and $\hat{p}$ to the Fock vacuum (\ref{state with gamma}), where $\hat{p}$ may be replaced by the appropriate number. Since the Fock vacuum is annihilated by the antighost modes $\rb$ and $\rb_{MN}$ and also these modes are BRST trivial (\ref{BRST triviallity of generator}), we merely consider the subspace satisfying the conditions 
\begin{eqnarray}
      {\cal H} |\Psi \rangle = {\cal R}_{MN} |\Psi \rangle =0, \quad   
      \rb |\Psi \rangle = \rb_{MN} |\Psi \rangle =0 .
         \label{subspace of physical state}
\end{eqnarray}
On this subspace, the BRST-invariant state coincides with the $\hat{Q}$-invariant state.

For the time being, we analyze physical states in the subspace (\ref{subspace of physical state}) described in the following form:
\begin{equation}
    |\Psi \rangle = {\cal O} \left( \hat{p}, a^\dag_{JM}, b^\dag_{JM}, c^\dag_{J(Mx)}, d^\dag_{J(Mx)}, e^\dag_{J(My)}  \right) 
                     |\gm \rangle .
      \label{restricted state of psi}
\end{equation}
The operator ${\cal O}$ and the Riegert charge $\gm$ will be determined from the BRST invariance condition below. The state that ${\cal O}$ includes creation modes of gauge ghosts and antighosts is discussed later.

Since $\rc_M |\Psi \rangle =0$ for the state (\ref{restricted state of psi}), the $\hat{Q}$-invariance condition is expressed as 
\begin{eqnarray*}
    \hat{Q} |\Psi \rangle = \sum_M \rc^\dag_M Q_M |\Psi \rangle =0 .
\end{eqnarray*}
Thus, together with the Hamiltonian and rotation invariance conditions in (\ref{subspace of physical state}), we reproduce the physical state conditions
\begin{eqnarray}
   (H-4) |\Psi \rangle = R_{MN} |\Psi \rangle = Q_M |\Psi \rangle =0
       \label{physical state condition}
\end{eqnarray}
studied in \cite{hh, hamada05, hamada09, hamada12b}. Here, the condition for $Q_M^\dag$ is not necessary. This shows that the state $|\Psi \rangle$ is given by a primary scalar state with conformal weight $4$.\footnote{ 
The primary state is, in general, defined by $H |h, \{ r \} \rangle = h |h, \{ r \} \rangle$, $R_{MN} |h, \{ r \} \rangle = \left( \Sigma_{MN} \right)_{\{ r^\pp \},\{ r \}} |h, \{ r^\pp \} \rangle$ and $Q_M |h, \{ r \} \rangle =0$, where $h$ is the conformal weight, $\{ r \}$ denotes a representation of $SU(2)\times SU(2)$, and $\Sigma_{MN}$ is the generator of spin rotations of the state. The descendant state is generated by applying $Q_M^\dag$ to the primary state $|h, \{ r \} \rangle$.
} 

\subsection{Gravitational Primary States}

First of all, in order to find the operator ${\cal O}$ satisfying the conditions (\ref{physical state condition}), we look for creation operators that commute with the generator $Q_M$. The operator ${\cal O}$ will be constructed by combining them in a rotation-invariant form.

We first consider the Riegert sector. The commutation relations between $Q_M$ and the creation modes are given by
\begin{eqnarray*}
    && \left[ Q_M, \hat{q} \right] = -a_{\half M}, \qquad  \left[ Q_M, \hat{p} \right] = 0 , 
                  \nonumber \\
    && \left[ Q_M, a^\dag_{\half M_1} \right] = \left( \sq{2b_c}-i\hat{p} \right) \dl_{M M_1}  ,
               \nonumber  \\
    && \left[ Q_M, a^\dag_{JM_1} \right] =  \a \left( J-\half \right) \sum_{M_2} 
         \C^{\half M}_{J M_1, J-\half M_2} \eps_{M_2} a^\dag_{J-\half -M_2} 
\end{eqnarray*}
for $J \geq 1$ and 
\begin{eqnarray}
    && \left[ Q_M, b^\dag_{JM_1} \right] =
        - \sum_{M_2} \C^{\half M}_{J M_1, J+\half M_2} \eps_{M_2} a^\dag_{J+\half -M_2}
             \nonumber \\
    && \qquad\qquad\qquad\,\, 
        -\b \left( J-\half \right) \sum_{M_2}  \C^{\half M}_{J M_1, J-\half M_2} \eps_{M_2} b^\dag_{J-\half -M_2} 
        \label{transformation law of negative-metric mode}
\end{eqnarray}
for $J \geq 0$.

Since there is no creation mode that commutes with $Q_M$ for the Riegert sector, we consider creation operators constructed in a bilinear form. Such operators have been studied in Ref.\cite{hamada05} using the crossing properties of the $SU(2) \times SU(2)$ Clebsch-Gordan coefficients (\ref{crossing relations}).  We then find that there are two types of $Q_M$-invariant creation operators with conformal weight $2L$ for integers $L \geq 1$:
\begin{eqnarray*}
    S^\dag_{L N} &=& \fr{\sq{2}( \sq{2b_c}-i\hat{p} )}{\sq{(2L - 1)(2L + 1)}}  a^\dag_{L N}
            \nonumber  \\
    &&  + \sum_{K=\half}^{L - \half} \sum_{M_1} \sum_{M_2} x(L,K) 
        \C^{L N}_{L - K M_1, K M_2} a^{\dag}_{L - K M_1} a^{\dag}_{K M_2},
                \nonumber \\
    {\cal S}^\dag_{L - 1 N}  &=& -\sq{2} \left( \sq{2b_c} -i\hat{p} \right) b^\dag_{L - 1 N}
            \nonumber  \\
    &&  + \sum_{K=\half}^{L - \half} \sum_{M_1} \sum_{M_2} x(L,K) 
        \C^{L - 1 N}_{L - K M_1, K M_2} a^{\dag}_{L - K M_1} a^{\dag}_{K M_2}
            \nonumber  \\
    &&  + \sum_{K=\half}^{L-1} \sum_{M_1} \sum_{M_2} y(L,K)
         \C^{L - 1 N}_{L - K - 1 M_1, K M_2} b^\dag_{L - K - 1 M_1} a^{\dag}_{K M_2} ,
\end{eqnarray*}
where
\begin{eqnarray}
    x(L,K) &=&  \fr{(-1)^{2K}}{\sq{(2L-2K+1)(2K+1)}}\sq{ \left( \begin{array}{c}
                                     2L \\
                                     2K
                                     \end{array} \right)
                             \left(   \begin{array}{c}
                                     2L-2 \\
                                     2K-1
                                     \end{array} \right) },
                  \nonumber \\
      y(L,K)  &=&  -2\sq{(2L-2K-1)(2L-2K+1)}x(L,K)  .
         \label{x(L,K) and y(L,K) for building blocks of gravitational fields}
\end{eqnarray}

\begin{table}[h]
\begin{center}
\begin{tabular}{|c|c|}  \hline
  rank of tensor            & $0$   \\ \hline
  $Q_M$-inv. operators      & $S^\dag_{LN}$  \\
                            & ${\cal S}^\dag_{L-1N}$   \\ \hline
  weights  $(L \in {\bf Z}_{\geq 1})$   &  $2L $   \\ \hline
\end{tabular} 
\end{center}
\vspace{-3mm}
\caption{\label{table of building blocks for Riegert}{\small The building blocks of the gravitational primary states in the Riegert sector. Here, $L$ is an integer with $L \geq 1$.}}
\end{table}

The $Q_M$-invariant creation operator for the Weyl sector can be constructed in the same way. In the Weyl sector, there is a creation mode that commute with $Q_M$ (\ref{Q_M for Weyl sector}), which is the lowest positive-metric mode of the transverse-traceless field, $c^\dag_{1(Mx)}$, only. All other creation modes do not commute with $Q_M$ and so we need to find the $Q_M$-invariant creation operators constructed in a bilinear form. Such operators can also be found by using the crossing properties of $SU(2)\times SU(2)$ Clebsch-Gordan coefficients with tensor indices. They  have rather complicated forms with tensor indices up to rank 4, which are summarized in the right side of Table \ref{table of building blocks for Weyl}.

\begin{table}[h]
\begin{center}
\begin{tabular}{|c|ccccc|}  \hline
  rank of tensor  & $0$  & $1$ & $2$ &$3$ & $4$    \\ \hline
  $Q_M$-inv. operators &  $A^\dag_{LN}$ & $B^\dag_{L-\half(Ny)}$ & $c^\dag_{1 (Nx)}$ & $D^\dag_{L-\half (Nz)}$ & $E^\dag_{L(Nw)}$  \\
                 &   ${\cal A}^\dag_{L-1 N}$ &  &  &  & ${\cal E}^\dag_{L-1 (Nw)}$  \\ \hline
  weights $(L \in {\bf Z}_{\geq 3})$        &   $2L $  &   $2L $   &   $2$   &   $2L $   &   $2L $    \\ \hline
\end{tabular} 
\end{center}
\vspace{-3mm}
\caption{\label{table of building blocks for Weyl}{\small The building blocks of the gravitational primary states for the Weyl sector. $L$ is an integer with $L \geq 3$. The polarization indices $z, w$ are $\veps_3, \veps_4$, respectively. The explicit expressions of the buliding blocks for the Weyl section are given in Appendix.}}
\end{table}

By joining the operators summarized in Tables \ref{table of building blocks for Riegert} and \ref{table of building blocks for Weyl} using the $SU(2) \times SU(2)$ Clebsch-Gordan coefficients, we can construct the basis of primary states. Due to the crossing properties of the Clebsch-Gordan coefficients, any $Q_M$-invariant primary states can be conceivably deformed in such a fundamental form. Thus, the operators in Tables \ref{table of building blocks for Riegert} and \ref{table of building blocks for Weyl} provide the building blocks of primary states.

Using these building blocks, we can construct the gravitational primary states. For example, the primary states with conformal weight 2 are given by
\begin{eqnarray*}
     {\cal S}^\dag_{00} |\Om \rang,  \qquad  S^\dag_{1N} |\Om \rang,  \qquad  c^\dag_{1(Nx)}|\Om \rang .
\end{eqnarray*}
The first scalar state corresponds to the Ricci scalar $R$, apart from the conformal factor. The second tensor state with 9 components corresponds to the traceless part of the Ricci tensor $R_{\mu\nu} -g_{\mu\nu}R/4$. The last tensor state with 10 components corresponds to the Weyl tensor $C_{\mu\nu\lam\s}$, where the polarization $x=\pm 1$ stands for the self-dual and anti-self-dual components. The primary states with conformal weight 4 are given by three scalar states
\begin{eqnarray*}
      \left( {\cal S}^\dag_{00} \right)^2 |\Om \rangle,  \qquad
       \sum_{N} \eps_N S^\dag_{1-N}S^\dag_{1 N} |\Om \rang, \qquad
       \sum_{N,x} \eps_N c^\dag_{1(-Nx)}c^\dag_{1(Nx)}|\Om \rang  
\end{eqnarray*}
and two tensor states
\begin{eqnarray*}
     {\cal S}^\dag_{1N} |\Om \rang,   \qquad  
       \sum_{N_1, x_1} \sum_{N_2, x_2}\E^{1N}_{1(N_1 x_1),1(N_2 x_2)} c^\dag_{1(N_1 x_1)} c^\dag_{1(N_2 x_2)} |\Om \rang 
\end{eqnarray*}
The first three scalar states correspond to $R^2$, $(R_{\mu\nu}-g_{\mu\nu}R/4)^2$ and $C_{\mu\nu\lam\s}^2$, respectively. The fourth and fifth tensor states correspond to the traceless energy-momentum tensor for the Riegert and Weyl sectors, respectively.

Here, note that some of these primary states do not satisfy the unitarity bound discussed in ordinary CFT. It is a characteristic feature of higher-derivative theories. However, these states do not satisfy the BRST invariance condition yet and so they are not gauge-invariant states. The physical states are obtained by further imposing the Hamiltonian and rotation-invariance conditions for these states.

\subsection{Physical States}

The physical state $|\Psi \rangle$ (\ref{restricted state of psi}) is now written in the form 
\begin{eqnarray*}
     |\Psi \rang = {\cal O} ( S^\dag, {\cal S}^\dag, \cdots ) |\gm \rang ,
\end{eqnarray*}
where the dots denote the other building blocks. The BRST invariance condition (\ref{physical state condition}) requires that the operator ${\cal O}$ have to satisfy the algebra
\begin{eqnarray*}
     \left[ H, {\cal O} \right] = l {\cal O}, \quad  \left[ R_{MN}, {\cal O} \right] = 0, 
     \quad  \left[ Q_M, {\cal O} \right] = 0 ,
\end{eqnarray*}
where the first condition implies that ${\cal O}$ has the conformal weight $l (\geq 0)$. Since building blocks have even conformal weights, the weight $l$ for ${\cal O}$ is given by even integers, which corresponds to the number of derivatives that the corresponding physical gravitational field has. All tensor indices are contracted out in a $R_{MN}$-invariant way. By solving $h_\gm + l -4 = 0$ that comes from the Hamiltonian condition in (\ref{physical state condition}), the Riegert charge $\gm$ is determined to be
\begin{equation}
     \gm_l = 2b_c \left( 1 - \sq{1-\fr{4-l}{b_c}} \right).
       \label{Riegert charge gamma}
\end{equation}
Here, we choose the solution such that $\gm_l$ approaches the canonical value $4-l$ in the large-$b_c$ limit. Note that $\gm_l$ is a real number because of $b_c  >4$ independent of the matter field contents.

We here present some physical states with lower $l$ up to $6$. The lowest weight state is simply given by $|\gm_0 \rangle$, which corresponds to $\sq{-g}$. The second lowest state with $l=2$ is given by
\begin{eqnarray*}
    {\cal S}^\dag_{00} |\gm_2 \rang ,
\end{eqnarray*} 
which corresponds to the Ricci scalar curvature $\sq{-g}R$. For $l=4$, there are three physical states 
\begin{eqnarray*}
       \left( {\cal S}^\dag_{00} \right)^2 |\gm_4 \rangle,  \qquad
       \sum_{M} \eps_M S^\dag_{1-M}S^\dag_{1 M} |\gm_4 \rang, \qquad
       \sum_{M,x} \eps_M c^\dag_{1(-Mx)}c^\dag_{1(Mx)}|\gm_4 \rang .
\end{eqnarray*}
Here, note that $\gm_4=0$ from (\ref{Riegert charge gamma}). These states correspond to $\sq{-g}R^2$, $\sq{-g}(R_{\mu\nu}-g_{\mu\nu}R/4)^2$ and $\sq{-g}C_{\mu\nu\lam\s}^2$, respectively.

In addition to the states discussed above, there are physical states with gauge ghost and antighost creation modes $\rc^\dag_M$ and $\rb^\dag_M$. For $l=2$, for example, we obtain another BRST-invariant state, 
\begin{eqnarray*}
     \left\{  - \left( \hbox{$\sq{2b_1}$} -i\hat{p} \right)^2 \sum_M \eps_M \rb^\dag_{-M} \rc_M^\dag  
              + \hat{h} \sum_M \eps_M a^\dag_{\half -M} a^\dag_{\half M} 
                              \right\} |\gm_2 \rangle ,
\end{eqnarray*}
where $\hat{h}=\hat{p}^2/2 + b_c$. This state is, however, equivalent to the physical state given before up to the BRST trivial state as
\begin{eqnarray*}
    \fr{1}{2\sq{2}} {\cal S}^\dag_{00} |\gm_2 \rangle + Q_\BRST |\Upsilon \rangle ,
\end{eqnarray*}
where $|\Upsilon \rangle = ( \hbox{$\sq{2b_1}$} -i\hat{p} ) \sum_M \eps_M \rb_{-M}^\dag a_{\half M}^\dag |\gm_2 \rangle$, which satisfies the conditions ${\cal H} |\Upsilon \rangle = {\cal R}_{MN}|\Upsilon \rangle = \rb |\Upsilon \rangle = \rb_{MN} |\Upsilon \rangle =0$. In general, it seems that the physical state depending explicitly on gauge ghosts and antighosts reduces to the standard form (\ref{restricted state of psi}) up to the BRST trivial state.

Finally, we briefly mention the norm structure. The naive inner products between gauge ghost vacua and their Hermitian conjugates vanish as ${}_\gh\langle 0|0 \rangle_\gh = {}_\gh\langle 0|\prod \rc_M^\dag \prod \rc_M|0 \rangle_\gh=0$, which is easily confirmed by inserting the anticommutation relations $\{ \rb, \rc \} = 1$ and $\{ \rb_{M N}, \rc_{L K} \} = \dl_{M L} \dl_{N K} -\eps_M \eps_N \dl_{-M K} \dl_{-N L}$ into the relevant expressions. So, we normalize the gauge ghost sector by inserting the operator $\vartheta = i \rc \prod \rc_{MN}$ satisfying $\vartheta^\dag = \vartheta$ as ${}_\gh\langle 0|\prod \rc_M^\dag  \vartheta \prod \rc_M |0 \rangle_\gh =1$.

The dual state of the physical state is given by choosing another solution of the Hamiltonian condition. Since (\ref{definition of conformal weight h_a}) satisfies the duality relation $h_\gm = h_{4b_c-\gm}$, the dual state has the form $|{\tilde \Psi}\rang = {\cal O}|4b_c-\gm_l \rang$. Although this state has no classical counterpart and so has no physical meanings, we can define the norm structure such as $\lang {\tilde \Psi}| \vartheta |\Psi \rang= \lang 4b_c-\gm_l |\vartheta|\gm_l \rang =1$ because the total Riegert charges cancel out due to $\langle \Om| e^{4b_c\phi_0(0)} |\Om \rangle =1$.

\subsection{Physical Field Operators}

The BRST transformations of the $\phi$ field and the gauge ghost are given by
\begin{eqnarray*}
    i \left[ Q_\BRST, \phi \right] = c^\mu \hnb_\mu \phi + \fr{1}{4} \hnb_\mu c^\mu , \qquad
    i \left\{ Q_\BRST, c^\mu \right\} = c^\nu \hnb_\nu c^\mu .
\end{eqnarray*}
We here introduce the ghost function defined by
\begin{eqnarray}
     \om = \fr{1}{4!} \eps_{\mu\nu\lam\s}c^\mu c^\nu c^\lam c^\s .
         \label{ghost function}
\end{eqnarray}
This function transforms as 
\begin{eqnarray*}
    i \left[ Q_\BRST, \om \right] = c^\mu \hnb_\mu \om = -\om \hnb_\mu c^\mu ,
\end{eqnarray*} 
where $c^\mu \om =0$ are used in the second equality.

The simplest field operator that transforms as a primary scalar is given by
\begin{eqnarray*}
    V = \l: e^{\gm_0 \phi} \r: = e^{\gm_0 \phi_>} e^{\gm_0 \phi_0} e^{\gm_0 \phi_<} ,
\end{eqnarray*}
where $\phi_>$ and $\phi_<$ are the creation and annihilation parts of $\phi$, respectively. The exponential function of the zero-mode part $\phi_0$ can be written as $e^{\gm_0 \phi_0} = e^{\hat{q}\gm_0/\sq{2b_c}} e^{\eta\hat{p}\gm_0/\sq{2b_c}} e^{-i\eta\gm_0^2/4b_c}$. This field operator transforms under the BRST transformation as 
\begin{eqnarray*}
    i \left[ Q_\BRST , V \right] = c^\mu \hnb_\mu V + \fr{h_{\gm_0}}{4} \hnb_\mu c^\mu V,   
\end{eqnarray*}
where the conformal weight $h_{\gm_0}$ is defined by the equation (\ref{definition of conformal weight h_a}).

Using this transformation law, we can show that the product $\om V$ becomes BRST-invariant as
\begin{eqnarray*}
     i \left[ Q_\BRST , \om V \right] = \fr{1}{4} \left( h_{\gm_0} -4 \right) \om \hnb_\mu c^\mu V= 0 ,
\end{eqnarray*}
provided $h_{\gm_0} =4$ as before. The physical field $V$ corresponds to the cosmological constant operator that reduces to the classical form of $\sq{-g}$ when we take the large-$b_c$ limit such that $\gm_0 \to 4$.

The Ricci scalar operator is given by
\begin{eqnarray*}
     W &=& \l: e^{\gm_2 \phi} \left( \hnb^2 \phi + \fr{\gm_2}{h_{\gm_2}} \hnb_\mu \phi \hnb^\mu \phi 
                                    - \fr{h_{\gm_2}}{\gm_2} \right) \r:
                   \nonumber \\
          &=& W^1 + \fr{\gm_2}{h_{\gm_2}} W^2 - \fr{h_{\gm_2}}{\gm_2} U
\end{eqnarray*}
with $U = \l: e^{\gm_2 \phi} \r:$ and
\begin{eqnarray*}
   W^1 &=& \hnb^2 \phi_> U + U \hnb^2 \phi_<  ,
                \nonumber \\
   W^2 &=& - \fr{1}{4}\pd_\eta \phi_0 \pd_\eta \phi_0 U 
           - \half \pd_\eta \phi_0 U \pd_\eta \phi_0
           - \fr{1}{4} U \pd_\eta \phi_0 \pd_\eta \phi_0                
                \nonumber \\
        && - \pd_\eta \phi_0 \left( \pd_\eta \phi_> U + U \pd_\eta \phi_< \right) 
           - \left( \pd_\eta \phi_> U + U \pd_\eta \phi_< \right) \pd_\eta \phi_0
                 \nonumber \\
        && + \hnb_\mu \phi_> \hnb^\mu \phi_> U +2 \hnb_\mu \phi_> U \hnb^\mu \phi_< 
                + U \hnb_\mu \phi_< \hnb^\mu \phi_<  ,
\end{eqnarray*}
where $\hnb^2=\hnb_\mu \hnb^\mu$. This operator multiplied by $\om$ satisfies the BRST invariance condition 
\begin{eqnarray*}
     \left[ Q_\BRST, \om W \right] =0 ,
\end{eqnarray*}
provided $h_{\gm_2}=2$. In the large-$b_c$ limit such that $\gm_2 \to 2$ and $\gm_2/h_{\gm_2} \to 1$, the operator $W$ reduces to the classical form of the Ricci scalar $\sq{-g}R$ divided by $-6$.\footnote{ 
The classical form of the Ricci scalar curvature is given by $d^4 x \sq{-g}R= d\eta d\Om_3 e^{2\phi}(-6 \hnb^2 \phi -6 \hnb_\mu \phi \hnb^\mu \phi +6)$ on the $R \times S^3$ background with unit $S^3$.} 

The state-operator correspondence is in general given by $\lim_{\eta \to i\infty} e^{-4i\eta}O_\gm |\Om \rangle  = |O_\gm \rangle$ for the physical state with the Riegert charge $\gm$ and the corresponding physical field operator $O_\gm$ that becomes BRST invariant when multiplied by $\om$.

For the physical fields $V$ and $W$, for instance, we can obtain the physical states as follows:
\begin{eqnarray*}
    |V \rangle = \lim_{\eta \to i\infty}e^{-4i\eta}V |\Om \rangle 
    = \lim_{\eta \to i\infty} e^{i(-4+h_{\gm_0})\eta} e^{\gm_0 \phi_>}e^{\fr{\gm_0}{\sq{2b_c}}\hat{q}} |\Om \rangle
    = e^{\gm_0 \phi_0(0)} |\Om \rangle 
\end{eqnarray*}
and 
\begin{eqnarray*}
   |W \rangle &=& \lim_{\eta \to i\infty}e^{-4i\eta}W |\Om \rangle 
           \nonumber \\
    &=& \lim_{\eta \to i\infty} e^{i(-4+h_{\gm_2})\eta} \left\{ \hnb^2 \phi_> -2i \pd_\eta \phi_> 
               + \fr{\b}{h_{\gm_2}} \hnb_\mu \phi_> \hnb^\mu \phi_> \right\} 
                   e^{\gm_2 \phi_>}e^{\fr{\gm_2}{\sq{2b_c}}\hat{q}} |\Om \rangle
          \nonumber \\
    &=& -\fr{\gm_2}{2\sq{2}b_c}{\cal S}^\dag_{00}e^{\gm_2 \phi_0(0)} |\Om \rangle .
\end{eqnarray*}
These limits exist only when $h_{\gm_0}=4$ and $h_{\gm_2}=2$, respectively, as is required from the physical condition.

Since the most singular term of the gauge ghost function (\ref{ghost function}) at the limit $\eta \to i\infty$ behaves as $\om \propto  e^{-4i\eta}\prod_M \rc_M$, the BRST invariant physical state discussed before is given by the limit 
\begin{eqnarray*}
    |O_\gm \rangle \otimes \prod_M \rc_M |0 \rangle_\gh = \lim_{\eta \to i\infty}\om O_\gm |\Om \rangle \otimes |0 \rangle_\gh .
\end{eqnarray*}

\section{Summary and Its Physical Significance}
\setcounter{equation}{0}

Since the discovery of exact solution in 2D quantum gravity, it has become known that diffeomorphism invariance at the quantum level is stronger than that at the classical level. Indeed, in order to construct the closed quantum diffeomorphism algebra, we need to add the Wess-Zumino action associated with conformal anomalies to the original gravitational action $I$.

In this article, we have presented two quantum gravity theories with the BRST conformal invariance as a quantum diffeomorphism invariance in 2 and 4 dimensions (see also the book \cite{hamadabook}). Owing to the background-free property, we can study these theories employing any conformally-flat background without affecting the result. Thus, we can formulate quantum gravity as a conventional quantum field theory defined on the non-dynamical background. This is a significant advantage in quantizing gravity. We have here analyzed the physical state employing the cylindrical background $R \times S^{D-1}$. We then find that these two theories have a similar algebraic structure. In general, the BRST conformal symmetry will arise in even dimensional quantum gravity, because there are conformally invariant dimensionless gravitational actions \cite{ds, hamada01}, while it is not known in odd dimensions.

The dynamics of gravity in 4 dimensions is controlled by the following renormalizable quantum gravity action \cite{hamada02, hamada14b, hm16, hm17}: 
\begin{eqnarray*}
   I = \int d^4x \sq{-g} \left[ - \fr{1}{t^2} C_{\mu\nu\lam\s}^2 - b G_4 + \fr{m_\pl^2}{16\pi} R - \Lam + {\cal L}_{\rm matter}  \right] ,
\end{eqnarray*}
where the Einstein term and the cosmological term are added in order to discuss a spacetime evolution below, which were neglected in the text because we studied the dynamics far beyond the Planck mass scale. The dimensionless coupling constant $t$ is to handle the traceless tensor field in the perturbation theory, while $b$ is not an independent coupling, which is introduced to renormalize the UV divergence proportional to the Euler density. Accompanied with this divergence, the Riegert action $S_{\rm R}$ (\ref{Riegert action}) with the coefficient $b_c$ (\ref{coefficient of Riegert action}) is induced.

The BRST conformal symmetry then arises at the $t \to 0$ limit, which is realized at the high-energy limit due to the negativity of its beta function. We call this nature ``asymptotic background freedom", distinguished from the conventional asymptotic freedom.

In the mode-expansions of the gravitational fields in 4 dimensions, both the positive-metric mode and the negative-metric mode occur. They all are necessary ingredients to form the closed algebra of the BRST conformal symmetry \cite{hamada12a, hamada12b}. The BRST conformal symmetry fully mixes the positive-metric mode and the negative-metric mode through the transformation (\ref{transformation law of negative-metric mode}) so that all of these modes themselves are not gauge invariant. The physical field is given by a real primary composite scalar field. Furthermore, its descendants become BRST trivial.

Since the gravitational quantity is generally known to be a real field, it is quite natural that the physical field is given by a real field even at the quantum level. The unitarity requires that the field operator retains the property of real number even in correlation functions. It will be guaranteed by the positivity of both the Riegert action and Weyl action, because the path integral is properly defined for such actions bounded from below.\footnote{ 
It can be understood by considering the Wick-rotated Euclidean theory on the conformally-flat background. Since the physical field is written in terms of the fundamental fields, $\phi$ and $h_{\mu\nu}$, and each mode does not appear explicitly, the positivity of these actions is essential.} 

Finally, we mention the physical significance of background-metric independence. Of course, the current universe is not scale-invariant. This suggests that a spacetime phase transition from the scale-invariant quantum spacetime to the classical spacetime with a scale occurred in the very early universe. The existence of the dynamical scale separating these two phases is indicated from the asymptotic behavior of the gravitational coupling $t$, which is denoted by $\Lam_{\rm QG}$. This scale means that there is no concept of distance inside the region shorter than the correlation length $\xi_\Lam =1/\Lam_{\rm QG}$, because spacetime is fully fluctuated quantum mechanically there. In this sense, $\xi_\Lam$ denotes the ``minimal length" of spacetime we can measure. Thus, spacetime is practically quantized by $\xi_\Lam$ without discretizing it and excitations in quantum gravity would be given by the mass of order $\Lam_{\rm QG}$. If there is a stable quantum gravity excitation, it becomes a candidate for dark matter.

The inflationary scenario suggested by recent cosmological observations \cite{wmap, planck} indicates that the value of the dynamical scale is given by \cite{hy}
\begin{eqnarray}
    \Lam_{\rm QG} \simeq 10^{17}{\rm GeV}
      \label{value of Lambda}
\end{eqnarray}
below the Planck mass $m_\pl$. Inflation then ignites on the Planck mass scale. This is because about the Planck scale, the Einstein term becomes effective, but the coupling constant $t$ is still small enough due to $m_\pl \gg \Lam_{\rm QG}$. Thus, the spacetime configuration satisfying $C_{\mu\nu\lam\s}=0$ with the mass scale $m_\pl$ is realized, which is nothing but the de Sitter spacetime. Specifically, such an inflationary solution has been found in the coupled system of the induced Riegert action and the Einstein action.

\begin{figure}[h]
\begin{center}
\vspace{5mm}
\includegraphics[scale=0.6]{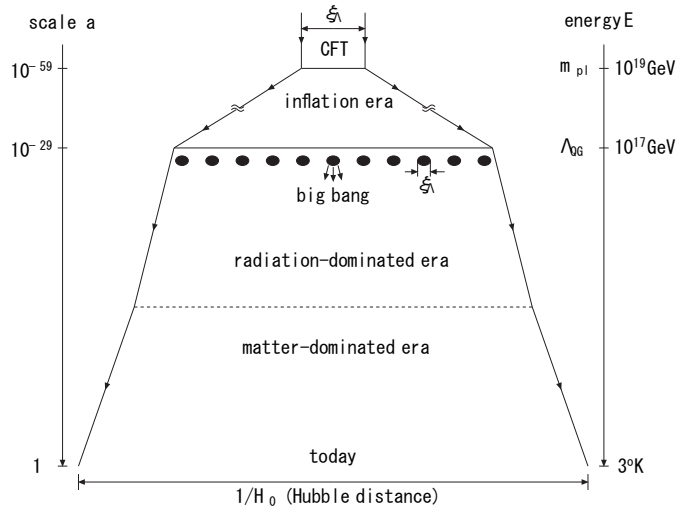}
\vspace{-3mm}
\end{center}
\caption{\label{evolution scenario} {\footnotesize Evolution scenario of the universe based on the quantum gravity cosmology \cite{hhy10}. The initial fluctuation prior to inflation with the correlation length $\xi_\Lam =1/\Lam_\QG ~(\gg l_\pl)$ is expanded up to the Hubble distance $1/H_0 (\simeq 5000{\rm Mpc})$ today. There is no correlation between two points with a distance larger than $\xi_\Lam$. Therefore, two points separated beyond $\xi_\Lam$ multiplied by the scale factor $10^{59}$ today lack correlation, and it is our resolution on the sharp fall-off of CMB angular power spectra at large angles.} }
\end{figure}

The evolution of the very early universe in the inflationary scenario driven by the quantum gravity effect (see Fig.\ref{evolution scenario}) is given as follows \cite{hy, hhy06, hhy10}. Below the Planck scale, quantum spacetime will start to fluctuate about the inflationary solution. The running coupling constant of $t$ increases during inflation. Along with that, the amplitude of the fluctuation gradually decreases. At the scale $\Lam_{\rm QG}$, the coupling diverges, and there the conformal invariance completely violates and eventually inflation terminates. Since the interactions causing the decay from the conformal-factor field to matter fields such as $\phi F_{\mu\nu}^2$ through conformal anomalies are opened about the transition point according to the increasing of the coupling, the spacetime transition like the big bang will occur, and then the classical spacetime in which particles are going to propagating emerges.

Physical observables can be estimated by the ratio of two mass scales as follows: the number of e-foldings is given by ${\cal N}_e \simeq m_\pl/\Lam_{\rm QG}$, the amplitude at the spacetime transition is estimated as $\dl R/R \sim (\Lam_{\rm QG}/m_\pl)^2$, where the denominator is given by the de Sitter curvature. This amplitude is consistent with the observed value when we take (\ref{value of Lambda}). The spectrum at the transition should reflect a conformal invariance of quantum gravity before inflation. That is almost scale-invariant and scalar-like, because physical states are given by scalars as shown in the text.  Furthermore, we can explain the sharp falloff of the low multipole components of CMB angular power spectrum in terms of the scale $\Lam_{\rm QG}$, because this inflationary scenario indicates that the universe has totally expanded until today about $10^{59}$ times during the inflationary epoch ($10^{30}$, ${\cal N}_e \simeq 70$) and subsequent expansion ($10^{29}$). It means that the present Hubble distance $1/H_0$ falls within the range of the correlation length $\xi_\Lam$ before inflation since the relationship $1/H_0 \simeq 10^{59} \xi_\Lam$ holds.

Below the energy scale $\Lam_{\rm QG}$, the Einstein action becomes dominated. Then, the decomposition of the metric field into the conformal-factor field and the traceless tensor field becomes inappropriate because of the loss of conformal symmetry. The dynamical field is, however, still given by the metric field as a composite field in which the conformal factor and the traceless tensor field are tightly coupled. The gravitational dynamics will be described by the low-energy effective theory given by an expansion in derivatives of the metric field about the Einstein action. The cosmological constant is then taken to be small,\footnote{More precisely, the physical cosmological constant that is defined as a renormalization group invariant quantity is taken to be small \cite{hm17}, and it will be passed on the low-energy effective theory of gravity.} 
which will explain the current accelerating universe.

\newpage

\appendix

\section{Building Blocks for Weyl Sector}

Here, we summarize the result for building blocks for tensor modes given in Table \ref{table of building blocks for Weyl} \cite{hamada05}. The commutation relations between $Q_M$ and tensor modes $c^\dag_{J(Mx)}$, $d^\dag_{J(Mx)}$ and $e^\dag_{J(My)}$ are given by
\begin{eqnarray*}
   \left[ Q_M, c^\dag_{J(M_1x_1)} \right] &=&
            \a\left( J-\half \right) \sum_{M_2,x_2} \eps_{M_2} \E^{\half M}_{J(M_1x_1),J-\half (-M_2x_2)}
            c^\dag_{J-\half (M_2x_2)} ,
               \nonumber \\ \\
   \left[ Q_M, d^\dag_{J(M_1x_1)} \right] &=&
         - \sum_{M_2,x_2} \eps_{M_2} \E^{\half M}_{J(M_1x_1),J+\half (-M_2x_2)}
            c^{\dag}_{J+\half (M_2x_2)}
                \nonumber  \\
        &&  -\b\left( J-\half \right) \sum_{M_2,x_2} \eps_{M_2} \E^{\half M}_{J(M_1x_1),J-\half (-M_2x_2)}
            d^\dag_{J-\half (M_2x_2)}
                \nonumber \\
        &&  -B(J)\sum_{M_2,y_2} \eps_{M_2} \H^{\half M}_{J(M_1x_1);J (-M_2y_2)}
            e^\dag_{J (M_2 y_2)} ,
                 \nonumber \\
   \left[ Q_M, e^\dag_{J(M_1y_1)} \right] &=&
           -A(J) \sum_{M_2,x_2} \eps_{M_2} \H^{\half M}_{J(-M_2x_2);J (M_1y_1)}
            c^{\dag}_{J (M_2 x_2)}
                \nonumber  \\
        &&  -C\left( J-\half \right) \sum_{M_2,y_2} \eps_{M_2} \D^{\half M}_{J(M_1y_1),J-\half (-M_2y_2)}
            e^\dag_{J-\half (M_2 y_2)} .
\end{eqnarray*}
From this, we can find that the creation mode that commutes with $Q_M$ is given by the lowest positive-metric mode $c^\dag_{1(Mx)}$ only. 
Furthermore, we can show that the $Q_M$ invariant building block with tensor index of rank 2 is given by this mode only, as shown in Table \ref{table of building blocks for Weyl}.

The building block with scalar index is given by the following two forms:
\begin{eqnarray*}
  && A^\dag_{L N} = \sum_{K=1}^{L-1} \sum_{M_1,x_1}\sum_{M_2x_2} x(L,K)
      \E^{LN}_{L-K (M_1x_1), K(M_2,x_2)} c^\dag_{L-K(M_1x_1)} c^\dag_{K(M_2x_2)} ,
             \nonumber \\
  && {\cal A}^\dag_{L-1 N} = \sum_{K=1}^{L-1} \sum_{M_1,x_1}\sum_{M_2x_2} x(L,K)
      \E^{L-1N}_{L-K (M_1x_1), K(M_2,x_2)} c^\dag_{L-K(M_1x_1)} c^\dag_{K(M_2x_2)}
               \nonumber \\
  && \quad 
     + \sum_{K=1}^{L-2} \sum_{M_1,x_1}\sum_{M_2x_2} y(L,K)
       \E^{L-1N}_{L-K-1 (M_1x_1), K(M_2,x_2)} d^\dag_{L-K-1 (M_1x_1)} c^\dag_{K(M_2x_2)}
               \nonumber  \\
  && \quad
     + \sum_{K=1}^{L-\fr{3}{2}} \sum_{M_1,x_1}\sum_{M_2y_2} w(L,K)
      \H^{L-1N}_{L-K-\half (M_1x_1); K(M_2,y_2)} c^\dag_{L-K-\half (M_1x_1)} e^\dag_{K(M_2y_2)}
               \nonumber \\
  && \quad
     +\sum_{K=1}^{L-2} \sum_{M_1,y_1}\sum_{M_2y_2} v(L,K)
      \D^{L-1N}_{L-K-1 (M_1y_1), K(M_2y_2)} e^\dag_{L-K-1(M_1y_1)} e^\dag_{K(M_2y_2)} ,
\end{eqnarray*}
where $L$ is positive integer and $x(L,K)$ and $y(L,K)$ are given by (\ref{x(L,K) and y(L,K) for building blocks of gravitational fields}). The new coefficients $w$ and $v$ are defined by
\begin{eqnarray*}
      w(L,K) &=&  2\sq{2} \sq{ \fr{(2L-2K-1)(2L-2K+1)}{2K(2K-1)(2K+3)} } x(L,K),
           \nonumber \\
      v(L,K) &=& -\sq{ \fr{(2K-1)(2K+2)(2L-2K-3)(2L-2K)}{(2K+3)(2L-2K+1)} }
                   x \left( L,K+\half \right) .
\end{eqnarray*}
Here, there is no $L=1$ case and also the $L=2$ case is trivial because it is composed of the $Q_M$ invariant mode $c^\dag_{1(Mx)}$ only. Therefore, the bilinear operator with $L \geq 3$ are new $Q_M$-invariant creation operators. On the other hand, we can find that there are no building blocks with half-integer $L$.

Furthermore, there are the building blocks with tensor indices of rank 1, 3 and 4. In order to define these building blocks, we introduce new $SU(2) \times SU(2)$ Clebsch-Gordan coefficient with $n$-th tensor harmonics, ${}^n\E$ and ${}^n\H$. The coefficients with $n=2$ are defined by
\begin{eqnarray*}
   && {}^2\E^{J (Mx)}_{J_1(M_1x_1), J_2(M_2x_2)}
      = \sq{\V3} \int_{S^3} d\Om_3 Y^{ij*}_{J(Mx)} Y^k_{i ~J_1(M_1x_1)} Y_{jk J_2(M_2x_2)},
                  \nonumber  \\
   && {}^2\H^{J (Mx)}_{J_1(M_1x_1); J_2(M_2y_2)}
      = \sq{\V3} \int_{S^3} d\Om_3 Y^{ij*}_{J(Mx)} Y^k_{i~J_1(M_1x_1)} \hnb_{(j}Y_{k) J_2(M_2y_2)} .
\end{eqnarray*}
The coefficient with $n=1$, ${}^1\E^{J (My)}_{J_1(M_1x_1), J_2(M_2x_2)}$ and ${}^1\H^{J (My)}_{J_1(M_1x_1); J_2(M_2y_2)}$, are defined by replacing $Y^{ij}_{J (Mx)}$ with $\hnb^{(i} Y^{j)}_{J(My)}$ in the expressions above. The coefficients with $n=4$ are
\begin{eqnarray*}
   && {}^4\E^{J (Mw)}_{J_1(M_1x_1), J_2(M_2x_2)}
      = \sq{\V3} \int_{S^3} d\Om_3 Y^{ijkl*}_{J(Mw)}
            Y_{ij J_1(M_1x_1)} Y_{kl J_2(M_2x_2)},
                   \nonumber \\
   && {}^4\H^{J (Mw)}_{J_1(M_1x_1); J_2(M_2y_2)}
      = \sq{\V3} \int_{S^3} d\Om_3 Y^{ijkl*}_{J(Mw)}
         Y_{ij J_1(M_1x_1)} \hnb_{(k}Y_{l) J_2(M_2y_2)} .
\end{eqnarray*}
The coefficients with $n=3$, ${}^3\E^{J (Mz)}_{J_1(M_1x_1), J_2(M_2x_2)}$ and ${}^3\H^{J (Mz)}_{J_1(M_1x_1); J_2(M_2y_2)}$, are defined by replacing $Y^{ijkl}_{J (Mw)}$ with $\hnb^{(i} Y^{jkl)}_{J(Mz)}$.

The building block with indices of $n=1$ vector harmonics is given for integer $L \,(\geq 3)$ by 
\begin{eqnarray*}
  && B^\dag_{L-\half (Ny)} = \sum_{K=1}^{L-1} \sum_{M_1,x_1}\sum_{M_2,x_2} x(L,K) \,
       {}^1\E^{L-\half (Ny)}_{L-K (M_1x_1), K(M_2x_2)} c^\dag_{L-K(M_1x_1)} c^\dag_{K(M_2x_2)}
               \nonumber \\
  && \qquad 
      + \sum_{K=1}^{L-\fr{3}{2}} \sum_{M_1,x_1}\sum_{M_2,y_2} w(L,K) \,
       {}^1\H^{L-\half (Ny)}_{L-K-\half (M_1x_1); K(M_2y_2)} c^\dag_{L-K-\half (M_1x_1)} e^\dag_{K(M_2y_2)} .
\end{eqnarray*}

The building block with indices of rank 3 tensor is given for integer $L \,(\geq 3)$ by 
\begin{eqnarray*}
  && D^\dag_{L-\half (Nz)} = \sum_{K=1}^{L-1} \sum_{M_1,x_1}\sum_{M_2,x_2} x(L,K) \,
      {}^3\E^{L-\half (Nz)}_{L-K (M_1x_1), K(M_2 x_2)} c^\dag_{L-K(M_1x_1)} c^\dag_{K(M_2x_2)}
               \nonumber \\
  &&\qquad
     + \sum_{K=1}^{L-\fr{3}{2}} \sum_{M_1,x_1}\sum_{M_2,y_2} w(L,K) \,
      {}^3\H^{L-\half (Nz)}_{L-K-\half (M_1x_1); K(M_2y_2)} c^\dag_{L-K-\half (M_1x_1)} e^\dag_{K(M_2y_2)} .
\end{eqnarray*}

There are two types of the building blocks with indices of rank 4 tensor, which are given for integer $L \,(\geq 3)$ by
\begin{eqnarray*}
  && E^\dag_{L (Nw)} = \sum_{K=1}^{L-1} \sum_{M_1,x_1}\sum_{M_2,x_2} x(L,K) \,
       {}^4 \E^{L(Nw)}_{~L-K (M_1x_1), K(M_2x_2)} c^\dag_{L-K(M_1x_1)} c^\dag_{K(M_2x_2)} ,
            \nonumber \\
  && {\cal E}^\dag_{L-1 (Nw)} = \sum_{K=1}^{L-1} \sum_{M_1,x_1}\sum_{M_2,x_2} x(L,K) \,
       {}^4\E^{L-1(Nw)}_{L-K (M_1x_1), K(M_2x_2)} c^\dag_{L-K(M_1x_1)} c^\dag_{K(M_2x_2)}
               \nonumber \\
   &&\qquad
     + \sum_{K=1}^{L-2} \sum_{M_1,x_1}\sum_{M_2,x_2} y(L,K) \,
       {}^4\E^{L-1(Nw)}_{L-K-1 (M_1x_1), K(M_2x_2)} d^\dag_{L-K-1 (M_1x_1)} c^\dag_{K(M_2x_2)}
               \nonumber  \\
   &&\qquad
     + \sum_{K=1}^{L-\fr{3}{2}} \sum_{M_1,x_1}\sum_{M_2,y_2} w(L,K) \,
       {}^4\H^{L-1(Nw)}_{L-K-\half (M_1x_1); K(M_2y_2)} c^\dag_{L-K-\half (M_1x_1)} e^\dag_{K(M_2y_2)} .
\end{eqnarray*}

\newpage



\begin{thebibliography}{99}

\bibitem{stelle}
K. Stelle, Phys. Rev. D {\bf 16} (1977) 953; Gen. Relativ. Gravit. {\bf 9} (1978) 353.  
\bibitem{tomboulis}
E. Tomboulis, Phys. Lett. {\bf 70B} (1977) 361; Phys. Lett. {\bf 97B} (1980) 77. 
\bibitem{ft}
E. Fradkin and A. Tseytlin, Nucl. Phys. {\bf B201} (1982) 469.  
\bibitem{ab}
I. Avramidy and A. Barvinsky, Phys. Lett. {\bf 159B} (1985) 269.
\bibitem{lw}
T. Lee and G. Wick, Nucl. Phys. {\bf B9} (1969) 209.
\bibitem{nakanishi}
N. Nakanishi, Prog. Theor. Phys. Suppl. {\bf 51} (1972) 1.
\bibitem{bg}
D. Boulware and D. Gross, Nucl. Phys. {\bf B233} (1984) 1. 
\bibitem{hs}
K. Hamada and F. Sugino, Nucl. Phys. {\bf B553} (1999) 283. 
\bibitem{hamada02}
K. Hamada, Prog. Theor. Phys. {\bf 108} (2002) 399.   
\bibitem{hamada14b}
K. Hamada, Phys. Rev. D {\bf 90} (2014) 084038.   
\bibitem{hm16}
K. Hamada and M. Matsuda, Phys. Rev. D {\bf 93} (2016) 064051.
\bibitem{hm17}
K. Hamada and M. Matsuda, Phys. Rev. D {\bf 96} (2017) 026010.
\bibitem{riegert}
R. Riegert, Phys. Lett. {\bf 134B} (1984) 56.
\bibitem{am}  
I. Antoniadis and E. Mottola, Phys. Rev. D {\bf 45} (1992) 2013.
\bibitem{amm92}
I. Antoniadis, P. Mazur and E. Mottola, Nucl. Phys. {\bf B388} (1992) 627. 
\bibitem{amm97a}
I. Antoniadis, P. Mazur and E. Mottola, Phys. Rev. D {\bf 55} (1997) 4756.
\bibitem{amm97b}
I. Antoniadis, P. Mazur and E. Mottola, Phys. Rev. D {\bf 55} (1997) 4770.
\bibitem{hh}
K. Hamada and S. Horata, Prog. Theor. Phys. {\bf 110} (2003) 1169.
\bibitem{hamada05}
K. Hamada, Int. J. Mod. Phys. A {\bf 20} (2005) 5353.
\bibitem{hamada09}
K. Hamada, Int. J. Mod. Phys. A {\bf 24} (2009) 3073.
\bibitem{hamada12a}
K. Hamada, Phys. Rev. D {\bf 85} (2012) 024028; {\bf 85} (2012) 124036.
\bibitem{hamada12b}
K. Hamada, Phys. Rev. D {\bf 86} (2012) 124006.
\bibitem{polyakov}
A. Polyakov, Phys. Lett. {\bf 103B} (1981) 207.
\bibitem{kpz}
V. Knizhnik, A. Polyakov and A. Zamolodchikov, Mod. Phys. Lett. A {\bf 3} (1988) 819.
\bibitem{dk}
J. Distler and H. Kawai, Nucl. Phys. {\bf B321} (1989) 509; F. David, Mod. Phys. Lett. A {\bf 3} (1988) 1651.
\bibitem{seiberg}
N. Seiberg, Prog. Theor. Phys. Suppl. {\bf 102} (1990) 319.
\bibitem{gl}
M. Goulian and M. Li, Phys. Rev. Lett. {\bf 66} (1991) 2051.
\bibitem{teschner}
J. Teschner, Class. Quant. Grav. {\bf 18} (2001) R153.
\bibitem{lz}
B. Lian and G. Zuckermann, Phys. Lett. {\bf B254} (1991) 417; {\bf B266} (1991) 21.
\bibitem{bmp}
P. Bouwknegt, J. McCarthy and K. Pilch, Commun. Math. Phys. {\bf 145} (1992) 541.
\bibitem{wz}
J. Wess and B. Zumino, Phys. Lett. {\bf 37B} (1971) 95.
\bibitem{cd}
D. Capper and M. Duff, Nuovo Cimento Soc. Ital. Fis. A {\bf 23} (1974) 173.
\bibitem{ddi}
S. Deser, M. Duff and C. Isham, Nucl. Phys. {\bf B111} (1976) 45.
\bibitem{duff}
M. Duff, Nucl. Phys. {\bf B125} (1977) 334.
\bibitem{hathrellS}
S. Hathrell, Ann. Phys. (N.Y.) {\bf 139} (1982) 136.
\bibitem{hathrellQED}
S. Hathrell, Ann. Phys. (N.Y.) {\bf 142} (1982) 34. 
\bibitem{freeman}
M. Freeman, Ann. Phys. (N.Y.) {\bf 153} (1984) 339.
\bibitem{ds}
S. Deser and A. Schwinmmer, Phys. Lett. {\bf B309} (1993) 279.
\bibitem{hamada01}
K. Hamada, Prog. Theor. Phys. {\bf 105} (2001) 673.
\bibitem{hamada14a}
K. Hamada, Phys. Rev. D {\bf 89} (2014) 104063.   
\bibitem{kato}
M. Kato and K. Ogawa, Nucl. Phys. {\bf B212} (1983) 443.
\bibitem{fms}
D. Friedan, E. Martinec and S. Shenker, Nucl. Phys. {\bf B271} (1986) 93.
\bibitem{witten}
E. Witten, Nucl. Phys. {\bf B373} (1992) 187.
\bibitem{kp}
I. Klebanov and A. Polyakov, Mod. Phys. Lett. A {\bf 6} (1991) 1.
\bibitem{klebanov}
I. Klebanov, Mod. Phys. Lett. A {\bf 7} (1992) 723.
\bibitem{hamada94}
K. Hamada, Phys. Lett. {\bf B324} (1994) 141.
\bibitem{ro}
M. Rubin and C. Ord\'{o}\~{n}ez, J. Math. Phys. {\bf 25} (1984) 2888. 
\bibitem{vmk}
D. Varshalovich, A. Moskalev and V. Khersonskii, {\it Quantum Theory of Angular Momentum} (World Scientific, Singapore, 1988).
\bibitem{hamadabook}
K. Hamada, {\it Quantum Gravity and Cosmology \small{based on conformal field theory}} [in Japanese] (Pleiades Publishing, 2016).
\bibitem{wmap}
G. Hinshaw {\it et al.} (WMAP9 Collaboration), Astrophys. J. Suppl. Ser. {\bf 208} (2013) 19.
\bibitem{planck}
P. Ade {\it et al.} (Planck Collaboration), Astron. Astrophys. {\bf 571} (2014) A16.
\bibitem{hy}
K. Hamada and T. Yukawa, Mod. Phys. Lett. A {\bf 20} (2005) 509. 
\bibitem{hhy06}
K. Hamada, S. Horata, and T. Yukawa, Phys. Rev. D {\bf 74} (2006) 123502.
\bibitem{hhy10}
K. Hamada, S. Horata, and T. Yukawa, Phys. Rev. D {\bf 81} (2010) 083533.


\end{thebibliography}
\end{document}